\documentclass{article}
 \usepackage[a4paper,top=2cm,bottom=2cm,left=2cm,right=2cm]{geometry}
\usepackage{authblk}
  \usepackage{paralist}
  \usepackage{graphics}
  \usepackage{epsfig} 
\usepackage{graphicx}  
\usepackage{epstopdf}
 \usepackage[colorlinks=true]{hyperref}
\hypersetup{urlcolor=blue, citecolor=red}
\usepackage{csquotes}
\usepackage{cite}

\usepackage{amsmath} 
\usepackage{amsfonts}
\usepackage{amssymb}
\usepackage{amsthm}
\usepackage{mathrsfs}
\usepackage{stackrel,graphicx}
\usepackage{enumitem}
\usepackage{cases}
\usepackage{amsmath}
\usepackage{booktabs}
\usepackage{braket}
\usepackage[strict]{changepage}
\usepackage{subfigure}
\usepackage[small,justification=justified]{caption}
\usepackage{titlesec} 
\usepackage{multicol}
\usepackage{hyperref}

\usepackage{amssymb,amsmath,amsthm}
\usepackage{booktabs}
\usepackage{multirow}
\usepackage[table]{xcolor}
\usepackage{subfigure}
\usepackage[right]{lineno}
\usepackage[capitalise]{cleveref}

\newtheorem*{rem*}{Remark}
\theoremstyle{definition}

\newcommand{\eps}{\varepsilon}
\newcommand{\R}{\mathbb R}
\newcommand{\Sbb}{\mathbb S^{n-1}}

\providecommand{\keywords}[1]{\textbf{\textit{Keyworks---}} #1}

\newenvironment{sistem}
{\left\lbrace\begin{array}{@{}l@{}}}
{\end{array}\right.}

\begin{document}

\title{Mathematical modeling of glioma invasion and therapy approaches via kinetic theory of active particles}
\author[1]{Martina Conte\thanks{Corresponding author: \texttt{conte.martina.93@gmail.com}}}
\author[2]{Yvonne Dzierma}
\author[2]{Sven Knobe}
\author[3]{Christina Surulescu}
\affil[1]{\centerline{\small Department of Mathematical Sciences "G. L. Lagrange", Politecnico di Torino} \newline \centerline{\small Corso Duca degli Abruzzi 24 - 10129 Torino, Italy}}
\affil[2]{\centerline{\small Saarland University Medical Centre, Department of Radiotherapy and Radiation Oncology} \newline \centerline{\small Kirrberger Str. Geb. 6.5 - 66421 Homburg/Saar, Germany}}
\affil[3]{\centerline {\small Technische Universit\"at Kaiserslautern, Felix-Klein-Zentrum f\"ur Mathematik} \newline \centerline{\small Paul-Ehrlich-Str. 31 - 67663 Kaiserslautern, Germany}}
\setcounter{Maxaffil}{0}
\renewcommand\Affilfont{\itshape\small}
\maketitle

\begin{abstract}
We propose here a multiscale model for study the effect of combined therapies on glioma spread in the brain under the influence of vascularization. The model accounts for the interplay between the different components of the neoplasm and the healthy tissue and it investigates and compares various therapy approaches. Precisely, these involve radio- and chemotherapy in a concurrent or adjuvant manner together with anti-angiogenic therapy affecting the vascular component of the system. We assess tumor growth and spread on the basis of DTI data, which allows us to reconstruct a realistic brain geometry and tissue structure, and we apply our model to real glioma patient data. In this latter case, a space-dependent radiotherapy description is considered using data about the corresponding isodose curves.
\end{abstract}
 

\section{Introduction}\label{sec:intro}   
Glioma is the most prevalent, aggressive, and invasive subtype of primary brain cancer, characterized by rapid cell proliferation and great infiltration capacity. Its infiltrative spread is strongly related to very poor patient survival prognosis, considering the difficulty to properly assess tumor margins for effective treatment\cite{Berens1999,shapiro99}. The tumor evolution takes advantage of the patient-specific anisotropic brain structure, in particular white matter tracts and blood vessels\cite{Giese2003,giese-etal96,onishi2011}. The anisotropic fiber structure characterizing the brain tissue influences the direction of cell migration determining preferential paths along which the cells move using specific receptor-binding mechanisms. Thus, the relevance of these specific structures in the overall tumor evolution motivates the need of developing personalized treatment planning, which should include such characteristics. 

The first approach to treat newly diagnosed gliomas is usually a surgical resection of the neoplasm. However, a complete resection is often impossible\cite{stummer2009extent} because of the irregular tumor outer rims, which cannot be assessed with the common medical imaging techniques. Therefore, supporting treatments are typically used in combination with surgical resection.

From the mathematical viewpoint, several classes of models have been proposed during the last two decades with the aim of using modern  biomedical visualization techniques to help predicting tumor volumes (in particular, CTV=clinical target volume, PTV=planning target volume) for therapy planning. 

 A first class of models gives pure macroscopic descriptions of glioma evolution in terms of diffusion-reaction equations directly stated at the macroscopic level \cite{Jbabdi2005,swanson2011}, on which they take into account  tumor-microenvironment interactions.

More recent approaches, instead, have a multiscale character and obtain effective equations for the whole tumor on the macroscale from the description of microscopic and/or mesoscale dynamics. These models are often built within the framework of the kinetic theory of active particles (KTAP)\cite{bellomo2021life}. In general, this theory has been applied in various contexts for describing complex systems in life sciences: besides biomathematics\cite{hillen2006m5}, a far-from-exhaustive list includes, for instance, crowd dynamics\cite{bellomo2022towards,albi2019vehicular} and vehicular traffic\cite{albi2019vehicular}, promoted by the pioneering articles Refs ~\cite{klar1997enskog} and ~\cite{prigogine1971kinetic}. 

Models within this context provide a more accurate description of multicellular systems according to a multiscale approach. They rely on the kinetic-based definition of cell dynamics at the mesoscopic level, thereby starting from processes that happen on the scale of single cells. Then, with suitable and well-known asymptotic techniques based on diffusive or hydrodynamic scalings, macroscopic equations  are derived from such descriptions on lower scales. In the context of glioma modeling, several different systems have been proposed on the basis of this approach\cite{conte2020,conte_surulescu2020,CEKNSSW,Corbin2018,dietrich2020,Engwer2,EHS,Engwer,Hunt2016,knobe2021,PH13,Swan_2017,ZS22}. Among these, Refs. ~\cite{conte_surulescu2020,CEKNSSW} suggest ways to assess tumor growth and grading. 

Settings taking into account the interplay of glioma with vascularization and other components (e.g., acidity) of the peritumoral space have been proposed within both model classes\cite{conte_surulescu2020,dietrich2020,kumar2020flux,swanson2011}. Multiscale cell-tissue interactions with time- and orientation-dependent fibrous extracellular structures have been addressed in Refs. ~\cite{CEKNSSW,hillen2006m5,Lorenz2014}. Since the common biomedical imaging techniques are not able to properly discern between the tissue constituents (blood vessels, extracellular matrix, white matter tracts, etc.) with their different functions, the question about how to model these structures can lead to several options. One of these is to combine the mentioned approach involving mesoscopic tissue dynamics with meso-or macro descriptions of endothelial cell distributions accompanied by time- and space-dependent macroscopic tissue characterizations\cite{conte_surulescu2020,dietrich2020,kumar2020flux}. 

Finally, models that pay attention to various therapy approaches are relatively scarce (see\cite{Holdsworth_2012,Rockne2010} for pure macroscopic settings and see\cite{Hunt2016} for a KTAP approach). Thus, there is still a need for mathematical formulations which aim at describing and comparing personalized treatment schedules involving state-of-the-art methods, with the purpose of identifying the most effective ones. This paper aligns to this aim. 

 The basis for kinetic descriptions of glioma and endothelial cell (ECs) dynamics can be found in Refs. ~\cite{conte2020,conte_surulescu2020}. Here we extend those settings upon considering a more detailed description at the microscopic level. In fact, subcellular dynamics of tumor receptor binding to anisotropic brain tissue and blood vessels are taken into account together with the dynamics of vascular endothelial growth factor (VEGF) receptors located on the EC membrane. With respect to Ref. ~\cite{conte_surulescu2020}, we consider here a different coupling between endothelial and tumor cell dynamics at the mesoscopic level. Moreover, we mainly focus on the description and inclusion of different terms characterizing the effects of radio-, chemo-, and anti-angiogenic therapy. The upscaling of the corresponding kinetic transport PDEs for glioma and endothelial cells leads in the parabolic limit to a system of reaction-advection-diffusion equations. We also couple with the macroscopic descriptions of VEGFs, healthy, and necrotic tissue evolution.

 The subsequent content is organized as follows. In Section 2, we describe the model setup on the subcellular and mesoscopic scales and we deduce the corresponding macroscopic PDEs. In Section 3, we concretize the coefficient functions and parameters, while the numerical simulations are presented in Section 4. In particular, the last part of this section is dedicated to the application of the proposed approach to real glioma patient data. A brief discussion is proposed in Section 5. 

\section{Modeling}\label{model}
In this note, we propose a multiscale model within the KTAP framework for glioma development and spread. Precisely, we integrate three different levels of description of cell dynamics - subcellular, mesoscopic, and population level - using the classical tools and methods of kinetic theory. This allows to derive mathematical structures suitable for the description of large systems of interacting entities. Our main aim is to interconnect the dynamics of glioma cells, vasculature, and vascular endothelial growth factors. 

Firslty, glioma cells mainly spread according to the anisotropy of brain tissue or moving along the blood vessels therein\cite{ferrer2018,onishi2011}. In order to provide nutrient and oxygen supply to its growing cell population, the tumor produces and releases in the extracellular environment growth factors, such as VEGFs, that attract endothelial cells. The resulting angiogenesis boosts the growth of the neoplasm, leading to an abundantly vascularized and progressed tumor mass. 

Standard treatment usually comprises chemo- and radiotherapy, in a concurrent or an adjuvant manner. 
During treatment, both tumor and normal tissue are affected, although to a different degree. Due to impaired tumor-associated vascularization, a necrotic region is also formed within the tumor mass. Additionally, we consider here the effect of an anti-angiogenic drug (e.g., bevacizumab\cite{joiner2009}), who is actually aimed at reducing the affinity of VEGF receptors on ECs  toward their ligand released in the extracellular space by the tumor cells. This therapeutic approach affects the EC tactic motility by reducing their bias towards (increasing gradients of) VEGFs. Moreover, it impairs their proliferation, which, in turn, leads to degradation of glioma through diminished nutrient supply. 

Starting from the subcellular level of interaction between cells, VEGFs, and tissue, we set up the corresponding kinetic transport equations (KTEs) for glioma cells and endothelial cells. Then, we perform a (non-rigorous) parabolic limit to deduce the macroscopic system of reaction-advection-diffusion PDEs for the involved cell populations: glioma and endothelial cells. These equations are finally coupled to the evolution of VEGF concentration, healthy tissue, and necrotic matter, which are considered directly on the macroscopic scale.

\subsection{Subcellular level}\label{cell level}
On the microscopic scale, we describe the interaction of glioma cells with the surrounding tissue including extracellular matrix (ECM), neuron bundles, and other components, thereby also considering the blood vessels (described via EC density), and the interaction of ECs with VEGFs diffusing in the extracellular space. The exchange of information between cells and their environment happens through receptors located on the cell membranes.  

\noindent
Considering such interactions follows the idea employed in Refs.  ~\cite{KELKEL2012,Lorenz2014} to build a micro-meso model for tumor invasion with chemo- and haptotaxis and revisited in the articles\cite{conte2020,CEKNSSW,Corbin2018,dietrich2020,Engwer2,EHS,Engwer,Hunt2016,Kumar20,kumar2020flux}, where upscalings from the KTEs on mesoscale have been performed, leading to PDEs with cross diffusion and various kinds of taxis. In particular, this approach allows to include not only the single-cell dynamics that usually appears in the KTEs, but also the subcellular level processes that highly impact the overall tumor progression. 
 
\noindent 
We denote by $y_1(t)$ the amount of tumor receptors bound to the blood vessels, by $y_2(t)$ the amount of tumor receptors bound to the tissue,  and by $\zeta(t)$ the amount of EC receptors bound to VEGF molecules. The corresponding binding dynamics are characterized by simple mass action kinetics: 
\begin{equation*}
\begin{split}
(\bar{R}_y-y_1-y_2)\,\,+\,\,W\stackrel[k_W^-]{k_W^+}{\rightleftharpoons}y_1\,,
\end{split}
\end{equation*}
\begin{equation*}
\begin{split}
(\bar{R}_y-y_1-y_2)\,\,+\,\,Q\stackrel[k_Q^-]{k_Q^+}{\rightleftharpoons}y_2\,,\\[0.2cm]
(\bar{R}_\zeta-\zeta)\,\,+\,\,H\stackrel[k_H^-]{k_H^+}{\rightleftharpoons}\zeta\,.
\end{split}
\end{equation*}
Here, $\bar{R}_y$ is the total amount of receptors on a tumor cell, and $\bar{R}_\zeta$ is that on an endothelial cell; both are assumed to be constant. We denote by $Q(t,x)$ the  macroscopic brain tissue density and by $W(t,x)$ the macroscopic density of ECs. They depend on the position $x \in \mathbb{R}^n$ and on the time $t>0$.  Further, $H(t,x)$ denotes the concentration of VEGFs. Accordingly, we get the ODE system
\begin{equation*}
\begin{split}
&\dot{y}_1=k_W^+\dfrac{W}{W_{c,0}}S_W(d_r,\alpha_W)\,(\bar{R}_y-y_1-y_2)-k_W^-y_1\,,\\[0.1cm]
&\dot{y}_2=k_Q^+\dfrac{Q}{Q^*}S_Q(d_r,\alpha_Q)(\bar{R}_y-y_1-y_2)-k_Q^-y_2\,,\\[0.1cm]
&\dot{\zeta}=k_H^+(d_b)\dfrac{H}{H_{c,0}}(\bar{R}_\zeta-\zeta)-k_H^-(d_b)\zeta ,
\end{split}
\end{equation*}
where $k_W^+$ and $k_W^-$, $k_Q^+$ and $k_Q^-$  represent attachment and detachment rates of glioma cells to vessels and tissue, respectively, while $k_H^+$ and $k_H^-$ are the corresponding rates in the process of  EC binding to VEGFs. The parameters $W_{c,0}$ and $Q^*$ are reference density values for endothelial cell and healthy tissue, respectively. At this subcellular level we also include the effect of therapy. In particular, we model the effect of radiotherapy on the normal tissue and on the ECs, with which glioma cells interact by way of their receptors. The function $S(d_r,\alpha):=exp(-\alpha d_r-\beta d_r^2)$ is used to model the fraction of cells that survive to a dose $d_r$ of radiotherapy: this is the so-called linear quadratic (LQ) formulation. Thereby, each cell population affected by radiotherapy has different values for the parameters $\alpha$ and $\beta$, referring to the lethal lesions produced by a single radiation track or by two radiation tracks\cite{fowler1989}. The relevant parameter in this model is the radiation sensitivity $\frac{\alpha}{\beta}$, which correlates to the cell cycle length: late responding tissues with a slow cell cycle have a small $\frac{\alpha}{\beta}$ ratio, while it is large for early responding, highly aggressive cancers. The parameters $\alpha_W$ and $\alpha_Q$ refer to the effect of radiotherapy on ECs and on brain tissue, respectively. In the following we will indicate with $S_W$ and $S_Q$ the radiotherapy survival fractions related to $W$ and $Q$, without explicitly mentioning their dependency on the parameters $\alpha$, $\beta$, and $d_r$. On the other hand, the interaction between ECs and VEGF is affected by the anti-angiogenic drug at dose $d_b$ that reduces the binding affinity of VEGF molecules to the corresponding receptors on ECs. In particular, $k_H^+:=\bar{k_H^+}l^+(d_b)$ and $k_H^-:=\bar{k_H^-}l^-(d_b)$, with $l^+(d_b)$ and $l^-(d_b)$ being a decreasing and an increasing function of $d_b$, respectively. \\
As in Ref. ~\cite{Engwer4}, we define $y:=y_1+y_2$ to be the total amount of transmembrane entities occupied by tissue or ECs, which allows us to lump together the two corresponding ODEs, into
\begin{equation*}
\dot{y}=\left(k_W^+\dfrac{S_W}{W_{c,0}}W+k_Q^+\dfrac{S_Q}{Q^*}Q\right)\bar{R}_y-y\left(k_W^+\dfrac{S_W}{W_{c,0}}W+k_Q^+\dfrac{S_Q}{Q^*}Q\right)-k_W^-y_1-k_Q^-y_2\,.
\end{equation*}

\noindent Assuming that $k_W^-=k_Q^-=k^-_1$, we get the microscopic equation for the subcellular dynamics of glioma cells
\begin{equation}
\dot{y}=\left(k_W^+\dfrac{S_W}{W_{c,0}}W+k_Q^+\dfrac{S_Q}{Q^*}Q\right)(\bar{R}_y-y)-k^-_1y\,.
\end{equation}

\noindent 
Rescaling $y/{\bar R}_y\leadsto y$ and $\zeta/{\bar R}_\zeta\leadsto \zeta$ will further simplify the notation. Then, the unique steady states of the above equations are given by:
\begin{equation*}
\begin{split}
&y^*=\dfrac{\left(k_W^+\frac{S_W}{W_{c,0}}W+k_Q^+\frac{S_Q}{Q^*}Q\right)}{k_W^+\frac{S_W}{W_{c,0}}W+k_Q^+Q\frac{S_Q}{Q^*}+k^-_1}=:\bar{f}(W,Q)\,,\\[0.1cm]
&\zeta^*=\dfrac{k_H^+\frac{H}{H_{c,0}}}{k_H^+(d_b)\frac{H}{H_{c,0}}+k_H^-(d_b)}=:\bar{e}(H)\,.
\end{split}
\end{equation*}
The variable $y$ characterizes the 'internal' state of tumor cells (called 'activity' in the kinetic theory of active particles, see\cite{bellomo2021life}). Likewise, $\zeta$ is the activity variable for ECs. In the sequel we will consider the mesoscopic densities $p(t,x,v,y)$ and $w(t,x,\vartheta,\zeta)$ of tumor cells and ECs, respectively, hence both depending on such activity variables. Thereby, $v\in V\subset \R^n$ and $\vartheta\in \Theta\subset\R^n$ are the velocity vectors of glioma and ECs, respectively, with the spaces $V$ and $\Theta$ to be closer explained in Subsection \ref{subsec:meso_level} below.

\noindent
For the glioma cells, we assume that they follow the gradients of tissue and vasculature, therefore we look at the path of a single cell starting at position $x_0$ and moving to position $x$ with velocity $v$ in the density fields $Q$ and $W$.
Denoting by $z:=y^*-y$ the deviation of $y$ from its steady state and introducing for simplicity the notation ${B_p(W,Q):=\left(k_W^+\dfrac{S_W}{W_{c,0}}W+k_Q^+\dfrac{S_Q}{Q^*}Q+k_1^-\right)}$, we have:
\begin{equation}
\begin{split}
\dot{z}=&\dfrac{k_1^-}{B_p(W,Q)^2}\! \left(k_W^+\dfrac{S_W}{W_{c,0}}\,v\!\cdot\! \nabla_x W+k_Q^+\dfrac{S_Q}{Q^*}\,v\!\cdot\! \nabla_x Q+\bar{F}_W(t)+\!\bar{F}_Q(t)\right)-zB_p(W,Q)\\[0.3cm]
&=:G(z,W,Q),
\end{split}
\end{equation}
with
\begin{equation*}
\begin{split}
&\bar{F}_W(t):=k_W^+\dfrac{S_W}{W_{c,0}}\partial_t W,\\[0.2cm]
&\bar{F}_Q(t):=k_Q^+\dfrac{S_Q}{Q_{c,0}}\partial_t Q\,.
\end{split}
\end{equation*}

With an analogous argument, denoting by $u=\zeta^*-\zeta$ the deviation of $\zeta$ from its steady state, we assume that the ECs follow the gradient of VEGF. Therefore, we look at the path of a single cell starting at position $x_0$ and moving to position $x$ with velocity $\vartheta$ in the concentration field $H$ so that $H(t,x)=H(t,x_0+\vartheta t)$.

\noindent
Thus, the equation for $u$ is given by:
\begin{equation}
\dot{u}=\dfrac{\frac{k_H^+}{H_{c,0}}k_H^-}{\left(\frac{k_H^+}{H_{c,0}}H+k_H^-\right)^2}\left(\partial_t\,H+\vartheta \cdot \nabla_x H\right)-u\left(\dfrac{k_H^+}{H_{c,0}}H+k_H^-\right)=:\Gamma(u,H)\,.
\end{equation}

For simplicity we denote $B_w(H):=\left(\dfrac{k_H^+}{H_{c,0}}H+k_H^-\right)$.

\subsection{Mesoscopic level}\label{subsec:meso_level}

We model the mesoscopic behavior of glioma and endothelial cells with the aid of KTEs describing velocity-jump processes and taking into account the subcellular dynamics. In particular, we follow a modified KTAP approach, including the introduced activity variable in a Fokker-Plank like framework and the microscale dynamics is modeled by an ODEs system rather than interactions. 

Concretely, we consider the following cell density functions:
\begin{itemize}
	\item $p(t,x,v,y)$ for glioma cells;
	\item $w(t,x,\vartheta,\zeta)$ for endothelial cells (ECs) forming capillaries, 
\end{itemize}
with the time and space variables $t>0$ and $x\in\R^n$, velocities $v\in V=s\mathbb{S}^{n-1}$ and $\vartheta\in \Theta :=\sigma \mathbb{S}^{n-1}$, and activity variables $y\in \mathcal{Y}=(0,1)$ and $\zeta \in \mathcal{Z}=(0,1)$. These choices mean that we assume for glioma and ECs constant speeds $s>0$ and $\sigma >0$, respectively, and we denote by $\mathbb{S}^{n-1}$ the unit sphere in $\mathbb{R}^n$. As in the articles\cite{Engwer2,EHS,Engwer,Hunt2016}, in the sequel we will work with the deviations $z=y^*-y\in \mathit{Y} \subseteq (y^*-1,y^*)$ rather then with $y$ for glioma cells, and with $u=\zeta^*-\zeta\in \mathit{U}\subseteq (\zeta^*-1,\zeta^*)$ rather then with $\zeta$ for ECs.
The corresponding macroscopic cell densities are denoted by $M(t,x)$ and $W(t,x)$, respectively.

\noindent
The kinetic transport equation for the glioma cell density function is given by:
\begin{equation}
\partial_t p +\nabla_x\cdot\,(v p) +\partial_z(G(z,W,Q,S_W,S_Q)p)=\mathcal{L}_p[\lambda(z)]p+\mathcal{P}\,p-(R_M(d_r)+C_M(k_c))p,
\label{peq}
\end{equation}
where $\mathcal{L}_p[\lambda(z)]p$ denotes the turning operator, describing tumor velocity changes. In particular, such changes are due to contact guidance: glioma cells tend to align according to the brain tissue anisotropy, mainly associated with white matter tracts. As in previous models of this type, $\mathcal{L}_p[\lambda(z)]p$ is a Boltzmann-like integral operator of the form
 \begin{equation} 
\mathcal{L}_p[\lambda(z)]p=-\lambda(z)p+\lambda(z)\int_V K(x,v)\,p(v')\,dv'\,,
 \label{turn_rate}
 \end{equation}
where $\lambda(z):=\lambda_0-\lambda_1z \ge 0$ is the cell turning rate  depending on the microscopic variable $z$, while $\lambda_0$ and $\lambda_1$ are positive constants. The integral term describes the reorientation of cells from any previous velocity $v'$ to a new velocity $v$ upon interacting with the tissue. The turning kernel $K(x,v)$ is assumed here to be independent on the incoming velocity $v'$ and it describes the dominating directional cue as given by the orientation of tissue fibers. Following Refs. ~\cite{Engwer2,PH13} we take $K(x,v):=\frac{q(x,\theta)}{\omega}$, where $\theta=\frac{v}{|v|}$ and $q(x,\theta)$, with $\theta \in \mathbb{S}^{n-1}$, is the orientational distribution of the fibers, normalized by $\omega =s^{n-1}$. This function encodes the information about the individual brain structure, obtained by diffusion tensor imaging (DTI). A concrete choice of $q(x,\theta)$ will be provided in Section \ref{CoefFun}. We also assume the tissue to be undirected (for a motivation refer to Ref. ~\cite{Kumar20}, where the model developed and its numerical simulations hint on the brain tissue being undirected), hence $q(x,\theta )=q(x,-\theta)$ for all $x\in \R^n$. For later reference we introduce the notations
\begin{align*}
&\mathbb E_q(x):=\int _{\Sbb} \theta q(x,\theta ) d\theta, \\
&\mathbb V_q(x):=\int _{\Sbb} (\theta -\mathbb E_q)\otimes (\theta -\mathbb E_q)\ q(x,\theta ) d\theta
\end{align*}
for the mean fiber orientation and the variance-covariance matrix for the orientation distribution of tissue fibers, respectively. Notice that the above symmetry of $q$ implies $\mathbb E_q=0$.

\noindent As in Refs. ~\cite{EHS,Hunt2016}, the term $\mathcal{P}\,p$ is used to describe the proliferation process triggered by cell receptor binding with the brain tissue and enhanced in the present formulation by the capillaries that provide the necessary nutrients and oxygen. We assume it to be given by:
\begin{equation*}
\mathcal{P}\,p:=\mu_M(M,W,N_e)\int_Z\chi(x,z,z^{'})\dfrac{Q(x)}{Q^*}p(t,x,v,z^{'})dz^{'}\,.
\end{equation*}
The proliferation rate $\mu(M,W,N_e)$, depends on the macroscopic densities of tumor cells and ECs ($M$ and $W$, respectively), but is also negatively influenced by necrotic matter, of density $N_e$. In the integral operator, the kernel $\chi (x,z,z')$ characterizes the transition from state $z'$ to state $z$ during such proliferation-initiating interaction at position $x$. No further conditions are required on $\chi$; we only assume that the nonlinear proliferative operator is uniformly bounded in the $L^2$-norm, which is reasonable because of space-limited cell division.

\noindent
The last terms in \eqref{peq} describe the decay of glioma cells due to radio- and chemotherapy. Thereby, $R_M(d_r)$ describes the radiation effect on tumor cells; its expression relates to the survival fraction $S(d_r,\alpha)$ as follows:
\begin{equation*}
R_M(d_r)=1-S(d_r,\alpha_M)\,,
\end{equation*}
with $d_r$ denoting the radiation dose and $\alpha_M$ being a parameter that refers (within the LQ model mentioned in \ref{cell level}) to the lesions caused by irradiation.  Chemotherapeutic effects inducing glioma depletion are quantified by $C_M(k_c)$, where the parameter $k_c$ represents the killing rate, related to the effectiveness of chemotherapy on the tumor cells. We consider that the same quantity of drug is administered once a day and that it has each time the same effectiveness, thus
\begin{equation}\label{dc}
	C_M(k_c):=k_c(t)=\sum _{i=1}^\nu k_c\eta_\delta (t-t_i),
\end{equation}
with $t_i$ denoting the time moments at which the drug is given and $\nu$ representing the number of times (days) allocated to this chemotherapy. Thereby, $\eta _\delta $ is a smooth function with unit mass and support in $(-\delta,\delta )$, with $\delta \ll 1$.
 
In particular, we indicate with $L_M(d_r,k_c):=(R_M(d_r)+C_M(k_c))$ the overall loss term for the tumor population.

\noindent
The KTE for ECs is given by:
\begin{equation}
\partial_t w+\nabla_x\cdot\,(\vartheta w)+\partial_u(\Gamma(u,H,d_b)w)=\mathcal{L}_w[\gamma(u)]w+\mathcal{P}_w\,w-R_W(d_r)w\,,
\label{weq}
\end{equation}
where the turning operator $\mathcal{L}_w[\gamma(u)]w$ describes changes in the orientation of ECs due to a chemotactic response to the concentration of growth factors produced by tumor cells. It is well-established, in fact, (e.g. see\cite{hanahan2011}) that tumor cells produce angiogenic signals acting as chemoattractants for ECs. These growth factors (mainly VEGF) stimulate endothelial cells to migrate toward the tumor and proliferate, a process called angiogenesis. Typically, the blood vessels produced within tumors and in their immediate proximity are distorted and enlarged, with erratic blood flux or microhemorrhages\cite{hanahan2011}. A switch from low to high vascularization is considered to be an essential indicator of tumor progression\cite{seyfried2012}.

\noindent
Similarly to the description of the turning operator for glioma cells, $\mathcal{L}_w[\gamma]w$ is an integral operator of the form
 \begin{equation} 
\mathcal{L}_w[\gamma(u)]w=-\gamma(u) w+\gamma(u) \int_{\Theta} \dfrac{1}{|\Theta|}w(\vartheta^{'})d\vartheta^{'}\,.
 \label{turn_rate}
 \end{equation}
In this case, for the turning kernel characterizing ECs reorientations we simply took a uniform distribution, while $\gamma(u):=\gamma_0-\gamma_1u \ge 0$, which represents the EC turning rate, depends on the microscopic variable $u$ and, consequently, on the concentration of VEGF, with $\gamma_0$ and $\gamma_1$ being some positive constants.

\noindent $\mathcal{P}_w\,w$ is the proliferation term related to ECs. It describes proliferation as a result of interactions between ECs and their chemoattractant VEGF
\begin{equation*}
\mathcal{P}_w\,w:=\mu_W(W,Q,d_b)\int_{U}\chi_W(x,u,u')\dfrac {H(t,x)}{H_{c,0}}w(t,x,\vartheta,u')du'\,.
\end{equation*}
The proliferation rate $\mu_W(W,Q,d_b)$ depends on the macroscopic density of ECs and on the density of brain tissue $Q$. We also consider that the anti-angiogenic therapy might also affect the proliferation of ECs more directly, in addition to the effect on the receptor binding kinetics. Thus, we include the dependency of the proliferation rate on the anti-angiogenic dose $d_b$.

\noindent Finally, $R_W(d_r)$ describes the depletion of ECs due to  radiotherapy. 
We take $R_W(d_r):=1-S(d_r,\alpha_W)$, with $\alpha_W>0$ a parameter related to the lesions caused on ECs by ionizing radiation.

\noindent
More details about the concrete choices of the coefficient functions involved in \eqref{peq} and \eqref{weq} will be provided in Section \ref{CoefFun}.

\subsection{Parabolic scaling of the mesoscopic model}\label{subsec:upscaling}
\noindent 
We deduce effective equations for the macroscopic dynamics of ECs and glioma cells, represented by the quantities $W$ and $M$, respectively. The PDEs for the VEGF concentration $H$, the density of healthy tissue $Q$ and of necrotic matter $N_e$ are stated directly at the macroscopic level. The heuristic derivation performed here is classical, see e.g. \cite{conte2020,conte_surulescu2020,CEKNSSW,Corbin2018,dietrich2020,Engwer2,EHS,Engwer,Hunt2016,Kumar20,kumar2020flux} for similar approaches in the same context. We also refer to \cite{burini2019multiscale} for a recent review attempting at a systematization and to \cite{ZS22} for a rigorous approach unifying parabolic and hyperbolic upscaling for a simplified equation. 

Concretely, we rescale here the time and space variables as $t\to\eps^2t$ and $x\to\eps x$ and we assume that the proliferation and death terms in \eqref{peq} and \eqref{weq} will be scaled by $\eps^2$ in order to account for the mitotic and apoptotic events taking place on a much larger time scale than migration. Hence, (\ref{peq}) and (\ref{weq}) become: 
\begin{equation}
\begin{split}
&\eps^2 \partial _tp+\,\eps \nabla _x\cdot (vp)-\partial_z (zB_p(W,Q)p)\\[0.2cm]
&+\partial_z \left(\left(\frac{k_1^-}{B_p(W,Q)^2}\left(\eps k_W^+\dfrac{S_W}{W_{c,0}}v\cdot \nabla_x W+\eps k_Q^+\dfrac{S_Q}{Q^*}v\cdot \nabla_x Q+\eps^2\left(\bar{F}_W+\bar{F}_Q\right)\right)\right)p\right)\\[0.2cm]
&=\mathcal{L}_p[\lambda(z)]p+\eps^2(\mathcal{P}-L_M(d_r,k_c))p\,,\label{peq-eps}
\end{split}
\end{equation}
\begin{equation}
\begin{split}
&\eps^2\partial_t w+\eps \nabla_x\cdot\,(\vartheta w) -\partial_u\Bigg(\Big(uB_w(H)-\frac{\frac{k_H^+}{H_{c,0}}k_H^-}{B_w(H)^2}\Big(\eps\vartheta\cdot \nabla_x H+\eps^2 \partial_t H\Big)\Big)w\Bigg)\\[0.2cm]
&=\mathcal{L}_w[\gamma(u)]w+\eps ^2( \mathcal P_w-R_W(d_r))w\,.
\end{split}
\label{weq-eps}
\end{equation}	
\noindent
As done in Refs. ~\cite{Engwer2,EHS,Engwer}, we define the following moments:
\begin{equation*}
\begin{split}
&m(t,x,v)=\int_Zp(t,x,v,z)dz,\quad \quad m^z(t,x,v)=\int_Z z\,p(t,x,v,z)dz,\\[0.2cm]
&M(t,x)=\int_Vm(t,x,v)dv,\quad\quad\quad M^z(t,x)=\int_Vm^z(t,x,v)dv,\\[0.2cm]
&\bar{w}(t,x,\vartheta)=\int_Uw(t,x,\vartheta,u)du,\quad \,\,\, \bar{w}^u(t,x,\vartheta)=\int_U u\, w(t,x,\vartheta,u)du,\\[0.2cm]
&W(t,x)=\int_\Theta \bar{w}(t,x,\vartheta)d\vartheta, \quad \,\quad\quad W^u(t,x)=\int_\Theta \bar{w}^u(t,x,\vartheta)d\vartheta,
\end{split}
\end{equation*}
and neglect the higher order moments w.r.t. the variables $z$ and $u$, in virtue of the subcellular dynamics being much faster than the events on the higher scales, hence $z \ll 1$ and $u \ll 1$. We assume the function $p$ to be compactly supported in the phase space $\mathbb{R}^n \times V \times \mathit{Z}$ and $w$ to be compactly supported in $\mathbb{R}^n \times \Theta \times \mathit{U}$.

\noindent
We first integrate equation (\ref{peq-eps}) with respect to $z$, getting the following equation for $m(t,x,v)$:
\begin{equation*}
\begin{split}
&\eps^2\partial_t m+\eps\,\nabla_x\cdot(v m)=-\lambda_0\left(m-\dfrac{q}{\omega}M\right)+\lambda_1\left(m^z-\dfrac{q}{\omega}M^z\right)\\[0.2cm]
&+\eps^2\int_Z\mu_M(M,W,N_e)\int_Z\chi(z,z',x)p(z')\dfrac{Q}{Q^*}dz'dz-\eps^2L_M(d_r,k_c)m\,.
\end{split}
\end{equation*}
Using the fact that $\chi(z,z',x)$ is a probability kernel with respect to $z$ for all $(x,z^{'})$, the previous equation for $m(t,x,v)$ reduces to:
\begin{equation}
\begin{split}
&\eps^2\partial_t m+\eps\nabla_x\cdot(v m)=-\lambda_0\left(m-\dfrac{q}{\omega}M\right)+\lambda_1\left(m^z-\dfrac{q}{\omega}M^z\right)\\[0.2cm]
&+\eps^2 \mu_M(M,W,N_e)\dfrac{Q}{Q^*}m-\eps^2 L_M(d_r,k_c)m.
\end{split}
\label{meq}
\end{equation}

\noindent Then, we multiply equation (\ref{peq-eps}) by $z$ and integrate it w.r.t. $z$, obtaining
{\begin{equation*}
\begin{split}
&\eps ^2\partial_t m^z+\eps \nabla_x\cdot(v m^z)-\int_Zz\partial_z \left[\left(zB_p(W,Q))-\eps^2\dfrac{k^-_1}{B_p(W,Q)^2}(\bar{F}_W+\bar{F}_Q)\right)p(z)\right]dz\\[0.2cm]
&\!\!+\int_Z \!\!z\partial_z\!\left[\eps \dfrac{k^-_1}{B_p(W,Q)^2}\! \left(k_W^+\dfrac{S_W}{W_{c,0}}v\!\cdot\!\nabla_x W\!+\!k_Q^+\dfrac{S_Q}{Q^*}v\!\cdot\!\nabla_x Q\!\right)p(z)\right]\!dz\!=\!\!\int_Z\!\!z\mathcal{L}_p[\lambda(z)]p(z)dz\\[0.2cm]
&\!+\!\eps^2\!\!\int_Z\!z\,\mu_M(M,W,N_e)\!\int_Z\!\chi(z,z',x)p(z')\dfrac{Q}{Q^*}dz'dz\!-\!\eps^2L_M(d_r,k_c)m^z.
\end{split}
\end{equation*}
The calculation of the integral term on the left hand side leads to the following equation for $m^z(t,x,v)$:
\begin{align}
&\eps^2\partial_t m^z\!\!+\!\eps\nabla_x\!\cdot\!(v m^z)\!=\dfrac{k^-_1}{B_p(W,Q)^2}\!\left(\eps\!\left(k_W^+\dfrac{S_W}{W_{c,0}}v\!\cdot\!\nabla_x W+k_Q^+\dfrac{S_Q}{Q^*}v\!\cdot\!\nabla_x Q\right)\!m\!\right)\notag \\[0.2cm]
&+\eps^2\mu_M(M,W,N_e)\int_Z\int_Z\,z\chi(z,z',x)p(z')\dfrac{Q}{Q^*}dz'dz-\eps^2L_M(d_r,k_c)m^z\notag \\[0.2cm]
&+\eps^2\dfrac{k^-_1}{B_p(W,Q)^2}\left(\bar{F}_W+\bar{F}_Q\right)m-\lambda_0\left(m^z-\dfrac{q}{\omega}M^z\right) -B_p(W,Q)m^z.
\label{mzeq}
\end{align}
Applying the same procedure to equation (\ref{weq-eps}), but integrating with respect to $u$, we obtain the following equation for the moment $\bar{w}$:
\begin{equation}
\begin{split}
\eps^2\partial_t \bar{w}+\eps\nabla_x\cdot(\vartheta \bar{w})=&-\gamma_0\left(\bar{w}\!-\!S_n^\sigma W\right)+\gamma_1\left(\bar{w}^u-S_n^\sigma W^u\right)\\[0.2cm]
&+\eps^2\mu_W(W,Q,d_b)\dfrac{H}{H_{c,0}}\,\bar{w}-\eps^2R_W(d_r)\bar{w}\,,
\end{split}
\label{barweq}
\end{equation}
assuming $\chi_W(x,u,u')$ to be a probability kernel with respect to $u$ for all $(x,u')$. Here we use the notation $S_n^\sigma:=\dfrac{1}{|\Theta |}=\dfrac{\sigma ^{1-n}}{|\mathbb S^{n-1}|}$.\\
Multiplying by $u$ and integrating (\ref{weq-eps}) again w.r.t. $u$, we obtain the equation for $\bar{w}^u$:
\begin{equation*}
\begin{split}
&\eps^2\partial_t \bar{w}^u\!+\!\eps\nabla_x\!\cdot\!(\vartheta \bar{w}^u)\!-\!\!\int_{U}\!\!u\partial_u\!\left[\!\left(\!uB_w(H)-\dfrac{\frac{k_H^+}{H_{c,0}}k_H^-}{B_w(H)^2}(\eps\vartheta\cdot\nabla_x H+\eps^2\partial_t H)\right)\!\!w(u)\!\right]\!du\\[0.2cm]
&=\int_{U}\!\!u\mathcal{L}_w[\gamma(u)]w(u)du+\eps^2\int_Uu\mu_W(W,Q,d_b)\int_U\chi_W(u,u',x)\dfrac{H}{H_{c,0}}w(u')du'du\\[0.3cm]
&-\eps^2R_W(d_r)\bar{w}^u\,.
\end{split}
\end{equation*}
After the calculation of the integral on the left hand side, it reduces to 
\begin{equation}\label{barwueq}
\begin{split}
&\eps^2\partial_t \bar{w}^u+\eps\nabla_x\cdot(\vartheta \bar{w}^u)+B_w(H)\bar{w}^u-\dfrac{\frac{k_H^+}{H_{c,0}}k_H^-}{B_w(H)^2}(\eps\vartheta\cdot\nabla_x H\,+\eps^2\partial_t H)\bar{w}\\[0.2cm]
&=\eps^2\mu_W(W,Q,d_b)\int_U\int_U\chi_W(u,u',x)\dfrac{H}{H_{c,0}}w(u')du'du\\[0.2cm]
&-\gamma_0\left(\bar{w}^u-S_n^\sigma W^z\right)-\eps^2R_W(d_r)\bar{w}^u\,.
\end{split}
\end{equation}
\noindent
We consider Hilbert expansions for the previously introduced moments:
\begin{equation*}
\begin{split}
&m(t,x,v)=\sum_{k=0}^{\infty}\eps^km_k,\quad   m^z(t,x,v)=\sum_{k=0}^{\infty}\eps^km^z_k,\quad  M(t,x)=\sum_{k=0}^{\infty}\eps^kM_k,\\[0.2cm]
&\bar{w}(x,t,\vartheta)=\sum_{k=0}^{\infty}\eps^k\bar{w}_k,\quad\bar{w}^u(x,t,\vartheta)=\sum_{k=0}^{\infty}\eps^k\bar{w}^u_k,\quad W(x,t)=\sum_{k=0}^{\infty}\eps^kW_k\,,\\[0.2cm]
& M^z(t,x)=\sum_{k=0}^{\infty}\eps^kM^z_k\,,\quad W^u(t,x)=\sum_{k=0}^{\infty}\eps^kW^u_k
\end{split}
\end{equation*}
For the subsequent calculations it will be useful to Taylor-expand the proliferation coefficient functions involving $W$ or $M$ in the scaled equations \eqref{meq}- \eqref{barwueq}:
\begin{equation*}
\begin{split}
&\begin{split}\mu (M,W,N_e)=&\mu (M_0,W_0,N_e)+\partial _M\mu (M_0,W_0,N_e)(M-M_0)\\[0.1cm]
&+\partial _W\mu (M_0,W_0,N_e)(W-W_0)+O(\eps ^2),
\end{split}\\[0.1cm]
&\mu_W(W,Q,d_b)=\mu_W(W_0,Q,d_b)+\partial_W\mu_W(W_0,Q,d_b)(W-W_0)+O(\eps ^2).
\end{split}
\end{equation*}
Moreover, we observe that
\[
B_p(W,Q)=B_p(W_0,Q)+ k_W^+\dfrac{S_W}{W_{c,0}}(W-W_0)+O(|W-W_0|^2) 
\]
and 
\begin{align*}
\dfrac{1}{(B_p(W,Q))^{2}}&\!=\!\dfrac{1}{(B_p(W_0,Q))^{2}}\!-\!\dfrac{2 k_W^+S_W}{W_{c,0}(B_p(W_0,Q))^{3}} (W-W_0)+O(|W-W_0|^2)\,.
\end{align*}

\noindent
Then, equating the powers of $\eps $ in the scaled equations \eqref{meq}-\eqref{barwueq}, we obtain:

\noindent $\eps^0$ terms:
\begin{align}
&0=-\lambda_0\left(m_0-\dfrac{q}{\omega}M_0\right)+\lambda_1\left(m_0^z-\dfrac{q}{\omega}M_0^z\right)\,,\label{m0eq}\\[0.2cm]
&0=-B_p(W_0,Q)m_0^z-\lambda_0\left(m_0^z-\dfrac{q}{\omega}M_0^z\right)\,, \label{m0zeq}\\[0.2cm]
&0=-\gamma_0(\bar{w}_0-S_n^\sigma W_0)+ \gamma_1(\bar{w}_0^u-S_n^\sigma W_0^u)\,, \label{w0eq}\\[0.2cm]
&0=-B_w(H)\bar{w}_0^u-\gamma_0\left(\bar{w}_0^u-S_n^\sigma W_0^u\right)\,. \label{w0ueq}
\end{align}

\noindent $\eps^1$ terms:
\begin{align}
&\nabla_x\cdot(v m_0)=-\lambda_0\left(m_1-\dfrac{q}{\omega}M_1\right)+\lambda_1\left(m_1^z-\dfrac{q}{\omega}M_1^z\right)\,, \label{m1eq}\\[0.5cm]
&\nabla_x\cdot(v m_0^z)+B_p(W_0,Q)m_1^z+k_W^+\dfrac{S_W}{W_{c,0}}W_1m_0^z \notag\\
&-\dfrac{k_1^-}{B_p(W_0,Q)^2}\,\left(k_W^+\dfrac{S_W}{W_{c,0}}v\cdot\nabla_x W_0+k_Q^+\dfrac{S_Q}{Q^*}v\cdot\nabla_x Q\right) m_0=-\lambda_0\left(m_1^z-\dfrac{q}{\omega}M_1^z\right)\,,\label{m1zeq} \\[0.5cm]
&\nabla_x \cdot(\vartheta \bar{w}_0)=-\gamma_0(\bar{w}_1-S_n^\sigma W_1)+\gamma_1(\bar{w}_1^u-S_n^\sigma W_1^u)\,, \label{w1eq}\\[0.5cm]
&\nabla_x\cdot(\vartheta \bar{w}_0^u)+B_w(H)\bar{w}_1^u-\dfrac{\frac{k_H^+}{H_{c,0}}k_H^-}{B_w(H)^2}\,\vartheta\cdot\nabla_x H \bar{w}_0=-\gamma_0\left(\bar{w}_1^u-S_n^\sigma W_1^u\right)\,.\label{w1ueq}
\end{align}

\noindent $\eps^2$ terms:
\begin{align}
&\partial_tm_0+\nabla_x\cdot(v m_1)=-\lambda_0\left(m_2-\dfrac{q}{\omega}M_2\right)+\lambda_1\left(m_2^z-\dfrac{q}{\omega}M_2^z\right)\notag\\
&+\mu_M(M_0,W_0,N_e)\dfrac{Q}{Q^*}\,m_0-L_M(d_r,k_c)m_0\,,\label{m2eq}\\[0.5cm]
&\partial_t \bar{w}_0+\nabla_x\cdot(\vartheta w_1)=-\gamma_0(\bar{w}_2+S_n^\sigma W_2)+\gamma_1(\bar{w}^u_2-S_n^\sigma W_2^u)\notag\\
&+\mu_W(W_0,Q,d_b)\dfrac{H}{H_{c,0}}\bar{w}_0-R_W(d_r)\bar{w}_0\,.\label{w2eq}
\end{align}

\noindent Integrating (\ref{m0zeq}) with respect to $v$, we have
\begin{equation*}
0=-M_0^zB_p(W_0,Q)\,,
\end{equation*}
i.e., 
\begin{equation}
M_0^z=0\,.
\label{m0zeq_1}
\end{equation}
\noindent
This, together with (\ref{m0zeq}), leads to 
\begin{equation}
m_0^z=0\,.
\label{m0zeq_2}
\end{equation}
From \eqref{m0eq} and \eqref{m0zeq} we obtain:
\begin{equation}
0=-\lambda_0m_0+\lambda_0\dfrac{q}{\omega}M_0\quad\Longrightarrow m_0=\dfrac{q}{\omega}M_0\,.
\label{m0eq_1}
\end{equation}
\noindent
From equation (\ref{w0ueq}) we have, by integrating w.r.t. $\vartheta$, 
\begin{equation*}
0=-W_0^uB_w(H)\,,
\end{equation*}
i.e.,
\begin{equation}
W_0^u=0\,.
\label{w0ueq_1}
\end{equation}
This, together with (\ref{w0ueq}), leads to 
\begin{equation}
\bar{w}_0^u=0\,.
\label{w0ueq_2}
\end{equation}
From \eqref{w0eq} and \eqref{w0ueq} we obtain:
\begin{equation}
0=-\gamma_0\bar{w}_0+\gamma_0S_n^\sigma W_0\quad\Longrightarrow \bar{w}_0=S_n^\sigma W_0\,,
\label{w0eq_1}
\end{equation}
thus $\bar{w}_0$ depends on the (constant) speed $\sigma$, but not on the direction $\vartheta\in \mathbb S^{n-1}$.\\

\noindent Now we turn to the equations stemming from the $\eps^1$-terms.
Integrating \eqref{w1ueq} w.r.t $\vartheta$ and using \eqref{w0eq_1} and \eqref{w0ueq_2}, we have
\begin{equation*}
0=-B_w(H)W_1^u+\dfrac{k_H^-\frac{k_H^+}{H_{c,0}}}{B_w(H)^2}\int_{\Theta} \vartheta\,\bar{w}_0\,\nabla_x H\,d\vartheta\,,
\end{equation*}
i.e.,
\begin{equation}
W_1^u=0
\label{w1ueq_1}
\end{equation}
due to \eqref{w0eq_1}. Plugging this result into \eqref{w1ueq} we get
\begin{equation}
\begin{split}
&0=-(B_w(H)+\gamma_0)\bar{w}_1^u+\dfrac{k_H^-\frac{k_H^+}{H_{c,0}}}{B_w(H)^2}\vartheta\cdot\nabla_x H\,\bar{w}_0\\[0.3cm]
&\Rightarrow \quad \bar{w}_1^u =\dfrac{k_H^-\frac{k_H^+}{H_{c,0}}}{(B_w(H)+\gamma_0)\,B_w(H)^2}\,\vartheta\cdot\nabla_x H \,\bar{w}_0\,.
\end{split}
\label{w1ueq_2}
\end{equation}
\noindent
Considering (\ref{w1eq}) and using (\ref{w1ueq_1}), we can derive
\begin{align}
&\nabla_x\cdot(\vartheta \bar{w}_0)=-\gamma_0\left(\bar{w}_1-S_n^\sigma W_1\right)+\gamma_1\bar{w}_1^u\nonumber\\[0.3cm]
&\Rightarrow  -\gamma_0 \bar{w}_1=\nabla_x\cdot(\vartheta \bar{w}_0)-\gamma_0S_n^\sigma W_1-\gamma_1\bar{w}_1^u\nonumber\\[0.3cm]
&\Rightarrow  \bar{\mathcal{L}}_W[\gamma_0]\bar{w}_1:=-\gamma_0 \bar{w}_1+\gamma_0S_n^\sigma W_1=\nabla_x\cdot(\vartheta \bar{w}_0)-\gamma_1\bar{w}_1^u\,.\label{pseudo-Lw-first}
\end{align}
In order to get an explicit expression for $w_1$, we would like to invert the operator $\bar{\mathcal{L}}_W[\gamma_0]$. Similarly to Ref. ~\cite{PH13} we define it on the weighted $L^2$-space $L^2_{S_n^\sigma}(\Theta)$, in which the measure $d\vartheta$ is weighted by $S_n^\sigma(\vartheta)$. In particular, $L^2_{S_n^\sigma}(\Theta)$ can be decomposed as $L^2_{S_n^\sigma}(\Theta)=<S_n^\sigma>\oplus<S_n^\sigma>^\perp$. Due to the properties of the chosen turning kernel, $\bar{\mathcal{L}}_W[\gamma_0]$ is a compact Hilbert-Schmidt operator with kernel $<S_n^\sigma>$. We can therefore calculate its pseudo-inverse on  $<S_n^\sigma>^\perp$. Thus, to determine $\bar{w}_1$ from \eqref{pseudo-Lw-first} we check the solvability condition, which holds for the above results on $\bar{w}_0$. Hence we obtain from \eqref{pseudo-Lw-first} and \eqref{w1ueq_2}
\begin{equation}
\bar{w}_1=-\dfrac{1}{\gamma_0}\left[\nabla_x\cdot(\vartheta \bar{w}_0)-\gamma_1\dfrac{\frac{k_H^+}{H_{c,0}}k_H^-}{(B_w(H)+\gamma_0)B_w(H)^2}\vartheta\cdot\nabla_x H\,\bar{w}_0\right]
\label{w1eq_1}
\end{equation}
and
\begin{equation}
W_1=0\,.
\label{w1eq_2}
\end{equation}
\noindent
On the other hand, integrating (\ref{m1zeq}) with respect to $v$, we have upon using \eqref{m0eq_1} and \eqref{m0zeq_2}:
\begin{equation*}
0=-B_p(W_0,Q)M_1^z+\dfrac{k^-_1}{B_p(W_0,Q)^2}M_0s^n\mathbb E_q\,\cdot\left(k_W^+\dfrac{S_W}{W_{c,0}}\nabla_x W_0+k_Q^+\dfrac{S_Q}{Q^*}\nabla_x Q\right)\,.
\end{equation*}
Using the fact that the tissue is undirected, i.e. $\mathbb E_q=0$, the previous equation leads to:
\begin{equation}
M_1^z=0\,.
\label{m1zeq_1}
\end{equation}
Now plugging these results into \eqref{m1zeq} we get
\begin{equation}
\begin{split}
&0=-(B_p(W_0,Q)+\lambda_0)m_1^z\!+\!\dfrac{k^-_1}{B_p(W_0,Q)^2} \left(k_W^+\dfrac{S_W}{W_{c,0}}v\cdot\nabla_x W_0+k_Q^+\dfrac{S_Q}{Q^*}v\cdot\nabla_x Q\right)m_0\\[0.3cm]
&\!\Rightarrow \,\, m_1^z =\dfrac{k^-_1}{(B_p(W_0,Q)+\lambda_0)\,B_p(W_0, Q)^2}\,\left(k_W^+\dfrac{S_W}{W_{c,0}}v\cdot\nabla_x W_0+k_Q^+\dfrac{S_Q}{Q^*}v\cdot\nabla_x Q\right)m_0.
\end{split}
\label{m1zeq_2}
\end{equation}
\noindent
Using (\ref{m1eq}) we derive
\begin{align}
&\nabla_x\cdot(v m_0)=-\lambda_0\left(m_1-\dfrac{q}{\omega}M_1\right)+\lambda_1m_1^z\nonumber\\[0.3cm]
&\Rightarrow  -\lambda_0 m_1=\nabla_x\cdot(v m_0)-\lambda_0\dfrac{q}{\omega}M_1-\lambda_1m_1^z\nonumber\\[0.3cm]
&\Rightarrow  \bar{\mathcal{L}}_M[\lambda_0]m_1:=-\lambda_0 m_1+\lambda_0\dfrac{q}{\omega}M_1=\nabla_x\cdot(v m_0)-\lambda_1m_1^z\,.\label{pseudo-L-first}
\end{align}
\noindent
As done for the turning operator of the endothelial cells, in order to get an explicit expression for $m_1$, we would like to invert the operator $\bar{\mathcal{L}}_M[\lambda_0]$. Again, similarly to Ref. ~\cite{PH13}, we define it on the weighted $L^2$-space $L^2_q(V)$, in which the measure $dv$ is weighted by $q(x,\theta)/\omega$. In particular, $L^2_q(V)$ can be decomposed as ${L^2_q(V)=<q/\omega>\oplus<q/\omega>^\perp}$. Due to the properties of the chosen turning kernel, $\bar{\mathcal{L}}_M[\lambda_0]$ is a compact Hilbert-Schmidt operator with kernel $<q/\omega>$. We can therefore calculate its pseudo-inverse on $<q/\omega>^\perp$. 

\noindent
Thus, to determine $m_1$ from \eqref{pseudo-L-first} we need to check the solvability condition
\begin{equation*}
\int_V \left[\nabla_x\cdot(v m_0)-\lambda_1m_1^z \right]dv=0\,.
\end{equation*}
This holds thanks to the above results and to the symmetry of $q(x,\theta)$. Therefore, we obtain from \eqref{pseudo-L-first} and \eqref{m1zeq_2}
\begin{equation}
\begin{split}
&m_1=-\dfrac{1}{\lambda_0}\nabla_x\cdot(v m_0)\\[0.2cm]
&+\dfrac{\lambda_1k^-_1}{\lambda_0(B_p(W_0,Q)+\lambda_0)B_p(W_0,Q)^2}\left(k_W^+\dfrac{S_W}{W_{c,0}}v\cdot\nabla_x W_0+k_Q^+\dfrac{S_Q}{Q^*}v\cdot\nabla_x Q\right)m_0
\end{split}
\end{equation}
and 
\begin{equation}
M_1=0\,.
\label{m1eq_1}
\end{equation}
\noindent
Integrating equation \eqref{m2eq} with respect to $v$ we get
\begin{equation}
\partial_tM_0+\int_V\nabla_x\cdot(v m_1)dv=\mu_M(M_0,W_0,N_e)\dfrac{Q(x)}{Q^*}M_0-L_M(d_r,k_c)M_0\,,
\label{m2eq_1}
\end{equation}
where
\begin{equation*}
\begin{split}
&\int_V\nabla_x\cdot(v m_1)dv= \int_V\nabla_x\cdot\Bigg[\,v\Bigg(-\dfrac{1}{\lambda_0}\bigg(\nabla_x\cdot(v m_0)\\[0.3cm]
&-\dfrac{\lambda_1k^-_1}{(B_p(W_0,Q)+\lambda_0)B_p(W_0,Q)^2}\left(k_W^+\dfrac{S_W}{W_{c,0}}v\cdot\nabla_x W_0+k_Q^+\dfrac{S_Q}{Q^*}v\cdot\nabla_x Q\right)\,m_0\bigg)\Bigg)\Bigg]dv \\[0.3cm]
&=\nabla_x\cdot\left[\int_V -\dfrac{1}{\lambda_0}v\otimes v\,\nabla_x\cdot\left(\dfrac{q}{\omega}M_0\right)\right]dv\\[0.3cm]
&+\nabla_x\cdot \dfrac{\lambda_1k_1^-k_W^+}{\omega \lambda_0(B_p(W_0,Q)+\lambda_0)B_p(W_0,Q)^2} \dfrac{S_W}{W_{c,0}}\nabla_x W_0 \int_V v\otimes v\,q(\theta,x)dv M_0\\[0.3cm]
&+\nabla_x\cdot \dfrac{\lambda_1k_1^-k_Q^+}{\omega \lambda_0(B_p(W_0,Q)+\lambda_0)B_p(W_0,Q)^2} \dfrac{S_Q}{Q^*}\nabla_x Q \int_V v\otimes v\,q(\theta,x)dv M_0.
\end{split}
\end{equation*}
With the notation
\begin{equation}
\mathbb D_T(x):=\dfrac{1}{\omega}\int_Vv\otimes v\,q(x,\theta)dv=s^2\int_{\mathbb{S}^{n-1}}\theta\otimes \theta\,q(x,\theta)d\theta=s^2\mathbb V_q(x)\,,
\label{DT}
\end{equation} 
we obtain from \eqref{m2eq_1} the following macroscopic equation for $M_0(t,x)$:
\begin{equation}
\begin{split}
&\partial_tM_0-\nabla_x \cdot \left[\dfrac{1}{\lambda_0}\nabla_x\cdot\left(\mathbb{D}_T M_0\right)\right]\\[0.3cm]
&+\nabla_x\cdot \left[\dfrac{\lambda_1k_1^-}{\lambda_0(B_p(W_0,Q)+\lambda_0)B_p(W_0, Q)^2}\mathbb{D}_T\left(k_W^+\dfrac{S_W}{W_{c,0}}\nabla_x W_0\,+k_Q^+\dfrac{S_Q}{Q^*}\nabla_x Q\right)M_0\right]\\[0.3cm]
&=\left(\mu_M(M_0,W_0,N_e)\dfrac{Q}{Q^*}-L_M(d_r,k_c)\right)M_0\,.
\end{split}
\label{M0_eq}
\end{equation}

\noindent
Integrating (\ref{w2eq}) with respect to $\vartheta \in \Theta$ gives
\begin{equation*}
\partial_t W_0+\int_{\Theta} \nabla_x\cdot(\vartheta \bar{w}_1) d\vartheta =\mu_W(W_0,Q,d_b)\dfrac{H}{H_{c,0}}W_0-R_W(d_r)W_0\,,
\end{equation*}
where
\begin{equation*}\hspace{-0.5cm}
\begin{split}
&\int_{\Theta} \nabla_x\!\cdot\!(\vartheta \bar{w}_1)d\vartheta\!=\!\int_{\Theta}\nabla_x \!\cdot\!\left[\vartheta\!\left(-\dfrac{\nabla_x\!\cdot\!(\vartheta \bar{w}_0)}{\gamma_0}\!+\!\dfrac{\gamma_1\frac{k_H^+}{H_{c,0}}k_H^-}{\gamma_0(B_w(H)+\gamma_0)B_w(H)^2}\vartheta\cdot\nabla_x H\,\bar{w}_0\right)\!\right]\\[0.3cm]
&=\nabla_x\cdot\left[\int_{\Theta} -\dfrac{1}{\gamma_0}\vartheta\otimes\vartheta\nabla_x\bar{w}_0\right]d\vartheta\\[0.3cm]
&+\nabla_x\cdot\Bigg[\dfrac{\gamma_1\frac{k_H^+}{H_{c,0}}k_H^-}{\gamma_0(B_w(H)+\gamma_0)B_w(H)^2} \int_{\Theta} \vartheta\otimes\vartheta\,d\vartheta\, \nabla_x H\,W_0\Bigg]\,.
\end{split}
\end{equation*}
\noindent
Recalling \eqref{w0eq_1}, this leads to the following macroscopic equation for the density $W_0$ of endothelial cells:
\begin{equation}
\begin{split}
&\partial_t W_0-\,\nabla_x\cdot\left(\mathbb D_{EC} \nabla_x W_0\right)+ \nabla_x\cdot\left(\mathbb D_{EC} \dfrac{\gamma_1k_H^+k_H^-}{H_{c,0}(B_w(H)+\gamma_0)B_w(H)^2}\nabla_x H \,W_0\right)=\\[0.2cm]
&\left(\mu_W(W_0,Q,d_b)\dfrac{H}{H_{c,0}}-R_W(d_r)\right)W_0\,,
\end{split}
\label{W0eq}
\end{equation}
where 
\begin{equation}
\mathbb D_{EC}:=\dfrac{\sigma^2}{n\gamma_0}\mathbb I_n\,.
\label{DEC}
\end{equation}
The two macroscopic equations obtained in \eqref{M0_eq} and \eqref{W0eq} for the evolution of glioma cells and ECs, respectively, are coupled with a PDE for the dynamics of VEGF concentration:
\begin{equation}
\partial_t H=D_H \Delta H +f(M_0,W_0,H)\,,
\label{eq:Seq}
\end{equation}
where $D_H>0$ is the constant VEGF diffusion coefficient, while $f(M_0,W_0,H)$ is a reaction term describing the processes of VEGF production by tumor cells in hypoxic situations and its uptake by ECs. The concrete form of the reaction term is defined in the next section.
The degradation of the healthy tissue is due to the acidity produced by tumor cells and to the effects of radiotherapy. Thus, we describe the evolution of healthy tissue by means of the following ODE:
\begin{equation}
\partial_t Q=-d_Q\dfrac{M_0}{M_{c,0}}Q-R_Q(d_r)Q\,,
\label{eq_Q}
\end{equation}
where $M_{c,0}$ denotes the reference density for tumor cells and $d_Q$ is the tissue degradation rate. The term $R_Q(d_r)Q$ collects the effects of the radiation on the healthy tissue and is described with the linear quadratic (LQ) model, which we previously introduced and used to formulate the survival fractions $S_W$ and $S_Q$. Finally, the ODE describing the evolution of necrotic matter  takes into account both tissue degradation and therapy effects on tumor, endothelial cells, and normal tissue, i.e,
\begin{equation}
\partial_t N_e=L_M(d_r,k_c)M+R_W(d_r)W_0+d_Q\dfrac{M_0}{M_{c,0}}Q+R_Q(d_r)Q\,.
\label{eq:Ne_eq}
\end{equation}

\noindent
In view of \eqref{m1eq_1} and \eqref{w1eq_2}, the $\eps$-correction terms for $M$ and $W$ can be left out and, neglecting the higher order moments, we get the following PDE system characterizing the macroscopic evolution of the tumor under the influence of tissue, vasculature, and growth factors:
\begin{equation}
\begin{sistem}
\partial_t M -\nabla_x \cdot \left[\dfrac{1}{\lambda_0}\nabla_x\cdot\left(\mathbb{D}_T M\right)\right]+\nabla_x\cdot\left[\dfrac{\lambda_1k_1^-}{\lambda_0(B_p+\lambda_0)B_p^2}\mathbb{D}_T\dfrac{S_Wk_W^+}{W_{c,0}}\nabla_x WM\right]\vspace{0.5cm}\\
+\nabla_x\cdot \left[\dfrac{\lambda_1k_1^-}{\lambda_0(B_p+\!\lambda_0)B_p^2}\mathbb{D}_T\dfrac{S_Qk_Q^+}{Q^*}\nabla_x QM\right]=\!\left(\!\mu_M(M,W,N_e)\dfrac{Q}{Q^*}\!-\!L_M(d_r,k_c)\!\right)\!M,\vspace{0.5cm}\\
\partial_t W-\,\nabla_x\cdot\left(\mathbb D_{EC} \nabla_x W\right)+ \nabla_x\cdot\Big(\mathbb D_{EC} \dfrac{\gamma_1k_H^+k_H^-}{H_{c,0}(B_w+\gamma_0)B_w^2}\nabla_x H \,W\Big)\vspace{0.5cm}\\
=\left(\mu_W(W,Q,d_b)\dfrac{H}{H_{c,0}}-R_W(d_r)\right)W
,\vspace{0.5cm}\\
\partial_t H=D_H \Delta H +f(M,W,H)\,,\vspace{0.5cm}\\
\partial_t Q=-d_Q\dfrac{M}{M_{c,0}}Q-R_Q(d_r)Q\,,\vspace{0.5cm}\\
\partial_t N_e=L_M(d_r,k_c)M+R_W(d_r)W+d_Q\dfrac{M}{M_{c,0}}Q+R_Q(d_r)Q\,,
\end{sistem}
\label{mac_set_neu}
\end{equation}
with the tumor diffusion tensor $\mathbb D_T$ given in \eqref{DT} and the EC diffusion tensor $\mathbb{D}_{EC}$ given by \eqref{DEC}. For simplicity of the notation, here we neglect the dependency of $B_p$ on $W$ and $Q$ and the dependency of $B_w$ on $H$. This system has been deduced considering $x\in \R^n$ and it has to be supplemented with adequate initial conditions. For the numerical simulations performed in the next Section \ref{simulation} we will consider the system to be set in a bounded, sufficiently regular domain $\Omega \subset \R^2$ and endow it with no-flux boundary conditions, which are obtained e.g., in a similar way to that presented in Ref. ~\cite{Plaza}.

\noindent
The next section is dedicated to specifying the remaining coefficients of the above macroscopic system.

\section{Assessment of coefficients}
\label{CoefFun}
To determine the tumor diffusion tensor $\mathbb D_T(x)$ in (\ref{DT}) we need to provide a concrete form for the (mesoscopic) orientational distribution of tissue fibers $q(x,\theta)$. Several different choices are available in the literature, e.g. see\cite{conte2020,PH13}. As in Ref. ~\cite{Hunt2016} we use the orientation distribution function (ODF)
\begin{equation}\label{qODF}
q(x,\theta)=\dfrac{1}{4\pi|\mathbb D_W(x)|^{\frac{1}{2}}(\theta^T(\mathbb D_W(x))^{-1}\theta)^{\frac{3}{2}}}\,,
\end{equation} 
where $\mathbb D_W(x)$ is the water diffusion tensor obtained from processing the (patient-specific) DTI data. 

\noindent
The growth rate $\mu(M,Q)$ can be also defined in different ways and it should be motivated by biological evidence. As in Ref. ~\cite{EHS}, we choose a logistic-like function to describe the growth self-limitation, enhancing this limitation with a factor depending on the necrotic tissue $Ne$
\begin{equation}
\mu(M,W,N_e)=\mu_{M,0}\,\left(1-\dfrac{M}{K_M}-\dfrac{N_e}{K_{Ne}}\right)g(W)\,.
\end{equation}
Here, $K_M$ is the tumor carrying capacity, $K_{Ne}$ is the carrying capacity of the necrotic matter, and $\mu_{M,0}$ is the proliferation rate. The function $g(W)$ describes the influence of vasculature on tumor growth. It can be simply a linear function of $W$, such as $g(W)=\frac{W}{W_{c,0}}$ describing tumor proliferation as an effect of interactions between tumor and ECs, or can be assumed to have an expression like $g(W)=(1+\frac{W}{W_{c,0}})$, describing ECs enhancing tumor proliferation in better-vascularized regions. For the simulations presented in Section \ref{simulation}, we take into account this second choice for the function $g(W)$.

\noindent
Similarly, for the term $\mu_W(W,Q,d_b)$ describing the proliferation rate of ECs, apart from a limited growth term, we have to include the influence of the healthy tissue on EC growth. In particular, the healthy tissue can be seen as necessary to drive ECs proliferation and, therefore, binary interactions between $W$ and $Q$ could be favorable
\begin{equation}
\mu_W(W,Q,d_b)=\mu_{W,0}(d_b)\left(1-\dfrac{W}{K_W}\right)\dfrac{Q}{Q^*}\,.
\end{equation}
Here, $Q^*$ is the reference density for the tissue, $K_W$ is the carrying capacity of the ECs, while $\mu_W(d_b)$ is the proliferation rate, depending on the dose of anti-angiogenic therapy $d_b$. This treatment, in fact, also impairs the ability of ECs to grow and, therefore, $\mu_W(d_b)$ can be described as a decreasing function of $d_b$.

\noindent 
Concerning radiotherapy, we already introduced the general expression related to the LQ model $S(d_r,\alpha):=\exp{(-\alpha d_r-\beta^2d_r)}$; precisely, the radiation-associated death for the two cell populations is described as
\begin{equation}
\begin{split}
&R_M(d_r):=\sum_{i=1}^\xi (1-S(d_r,\alpha_M))\eta_{\delta}(t-t_i),\\[0.3cm]
&R_W(d_r):=\sum_{i=1}^\xi (1-S(d_r,\alpha_W))\eta_{\delta}(t-t_i),
\label{radio-term}
\end{split}
\end{equation}
where the total dose $d_r$ of the drug is given in smaller fractions, to avoid toxic effects on healthy tissue. Thereby, $\xi$ is the number of fractions, $\eta_\delta$ is a $C_0^\infty$ function with unit mass and support in $(-\delta,\delta)$, $\delta\ll 1$, and $t_i$ denotes the time instants at which ionizing radiation is applied to the patient.
Moreover, the effects of anti-angiogenic therapy affecting ECs migration and proliferation are included in the binding/unbinding functions and in the proliferation term as
\begin{align}
&k_H^+:=\bar{k}_H^+l^+(d_b):=\bar{k}_H^+\left(\dfrac{1}{2}+\dfrac{1}{2(1+d_b^2)}\right)\,,\\[0.3cm]
&k_H^-:=\bar{k}_H^-l^-(d_b):=\bar{k}_H^-\left(1+d_b\right)\,,\\[0.3cm]
&\mu_{W,0}(d_b):=\bar{\mu}_{W,0}\left(\dfrac{1}{2}+\dfrac{1}{2(1+d_b^2)}\right)\,.
\end{align}

\noindent Finally, the reaction term in the PDE for the VEGF dynamics is chosen as 
\begin{equation}\label{concrete_f}
f(M_0,W_0,H):=c_1\frac{M_0}{M_{c,0}}\left(\dfrac{W_T}{W_{c,0}}-\dfrac{W}{W_{c,0}}\right)_+-c_2\,\frac{W_0}{W_{c,0}}H\,,
\end{equation}
where the parameters $c_1>0$ and $c_2>0$ represent the production and uptake rates of VEGFs, respectively, while $(\cdot)_+$ represents  the positive part. The parameter $W_T$ represents a threshold of EC density referring to hypoxic conditions. This means that when EC density is below such a threshold, the tumor cells keep producing VEGFs. The latter is produced in order to boost angiogenesis and therewith associated nutrient supply. The ECs are attracted by the VEGF gradient and uptake VEGF to sustain the growth of the capillary network.

\noindent
In Table 1 we report the (range of) values for the constant parameters involved in system (\ref{mac_set_neu}), as well as the references they were drawn from. In particular, concerning the parameters $\lambda_0$ and $k^+_{Q,S}$ we observe that the estimations are in good agreement with the ranges provided in Ref. ~\cite{knobe2021}, where the clinical applicability of a similar multiscale model in radiotherapy is studied for a collective of 3 glioma patients. Instead, following the results presented in Ref ~\cite{conte_surulescu2020}, we consider for the speed of tumor cells a smaller range of variability than that provided in Ref. ~\cite{knobe2021}.  
More details about the therapy are provided in the explanations of the corresponding simulations. 
\begin{table} [!h]\label{Tab_parameter}
\begin{center}
\caption{Model parameters (dimensional)}
 {\begin{tabular}{@{}c|c|c|c@{}}\toprule  
Parameter & Description & Value (unit) & Source \\
 \midrule
$\lambda_0$ & turning frequency in $\mathcal{L}_p[\lambda(z)]$ &$0.001$ (s$^{-1})$ & \cite{conte_surulescu2020,Sidani}\\
$\lambda_1$ & turning frequency in $\mathcal{L}_p[\lambda(z)]$ & $0.001$ (s$^{-1})$ &\cite{conte_surulescu2020,Sidani}\\
$s$ & tumor cells speed  & $[0.0042, 0.0084]\cdot10^{-3}$ (mm $\cdot$ s$^{-1})$ & \cite{conte_surulescu2020,diao2019}\\
$k^+_Q$ & tumor-ECM attachment rate & $0.034$ (s$^{-1})$&\cite{Lauffenburger,conte_surulescu2020}\\
$k^+_W$ & tumor-EC attachment rate & $0.034$ (s$^{-1})$&\cite{Lauffenburger}\\
$k^-$ & detachment rate & $0.01$ (s$^{-1})$ & \cite{Lauffenburger}\\
$\mu_{M,0}$ & tumor proliferation rate & $[3.5, 9]\cdot10^{-6}$ (s$^{-1})$ & \cite{ke2000}  \\
$K_M$ & tumor carrying capacity &  $\sim10^5$ (cells $\cdot$ mm$^{-3})$ &\cite{TCs_dim} \\
$K_{Ne}$ & necrotic matter carrying capacity &  $\sim10^5$ (cells $\cdot$ mm$^{-3})$&\cite{TCs_dim} \\
$K_{Q}$ & healthy tissue carrying capacity &  $\sim10^5$ (cells $\cdot$ mm$^{-3})$&\cite{swanson2011} \\
$\alpha_M$ & single radiation track lesion on tumor& $[0.04, 0.11]$ (Gy$^{-1})$ &\cite{qi2006,barazzuol2010}\\
$\alpha_Q$ & single radiation track lesion on tissue& $0.0025$ (Gy$^{-1})$& \cite{kroos2019}\\
$\alpha_W$ & single radiation track lesion on ECs & $0.0025$ (Gy$^{-1})$& this work\\
$\beta_M$ & two radiation tracks lesions on tumor  & $[0.006, 0.019]$ (Gy$^{-2})$ &\cite{qi2006,barazzuol2010}\\
$\beta_Q$ &  two radiation tracks lesions on tissue &$0.00005$ (Gy$^{-2})$ &\cite{kroos2019}\\
$\beta_W$ &  two radiation tracks lesions on ECs &$0.0008$ (Gy$^{-2})$&\cite{joiner2009}\\
$k_c$ & chemotherapy killing rate & $0.23\cdot 10^{-6}$ (s$^{-1})$& \cite{powathil2007}\\
$d_r$ &radiotherapy dose & $60$ (Gy) &this work\\
$\gamma_0$ &  turning frequency in $\mathcal{L}_w[\gamma(u)]$ & $2\cdot10^{-4}$ (s$^{-1})$&\cite{szabo2010,conte_surulescu2020}\\
$\gamma_1$ &  turning frequency in $\mathcal{L}_w[\gamma(u)]$ &$0.001$ (s$^{-1})$ &this work\\
$\sigma$ & ECs speed & $[0.0028, 0.0069]\cdot10^{-3}$ (mm $\cdot$ s$^{-1})$ &\cite{czirok2013,conte_surulescu2020}\\
$\bar{k}_H^+$ & EC-VEGF attachment rate & $0.03$ (s$^{-1})$& \cite{Popel2006} \\
$\bar{k}_H^-$ & EC-VEGF detachment rate &$0.001$ (s$^{-1})$& \cite{Popel2006} \\
$K_W$ & EC carrying capacity &  $\sim10^5$ (cells $\cdot$ mm$^{-3})$ &\cite{ECs_dim} \\
$\bar{\mu}_{W,0}$ & EC proliferation rate & $9.6\cdot10^{-6}$ (s$^{-1})$& \cite{conte_surulescu2020} \\
$W_T$ & EC threshold for hypoxia& $60\,\% \,K_W$ & this work \\
$d_b$ & anti-angiogenic therapy dose  & $10$ (mg $\cdot$ kg$^{-1})$&this work \\
$D_H$ & VEGF diffusion coefficient &$[10^{-8}, 10^{-7}]$ (mm$^2\cdot$ s$^{-1})$ & \cite{levine2001}  \\
$c_1$  & VEGF production rate & $1.4\cdot10^{-5}$ (s$^{-1})$ &  \cite{gevertz2006}\\
$c_2$ & VEGF consumption rate &$2.3\cdot10^{-4}$ (s$^{-1})$ &\cite{vempati2014}\\
$d_Q$ & tissue degradation rate (by acidity) &$0.58\cdot10^{-7}$ (s$^{-1})$ &\cite{conte_surulescu2020}\\[1ex]
\bottomrule
     \end{tabular}}
     \end{center}
\end{table}

\section{Numerical simulations}
\label{simulation}
We perform numerical simulations of the non-dimensionalized version of the macroscopic setting \eqref{mac_set_neu} (see Appendix \ref{adim_sis} for further details about the non-dimensionalization) in 2D with the parameters listed in Table 1. For the initial conditions we choose a Gaussian-like aggregate of tumor cells centered in $(x_{0,M},y_{0,M})=(-17,5)$, in the right hemisphere of the brain slice representing our illustrative computational domain $\Omega$ (in the following, $(x,y)\in\Omega\subseteq \mathbb{R}^2$):
\[
M_0(x,y)=e^{-\frac{(x-x_{0,M})^2+(y-y_{0,M})^2}{8}}\,.
\]
For the initial distribution of normal tissue we consider the expression proposed in Ref. ~\cite{EHS} 
\begin{equation}
Q_0(x,y)=1-\dfrac{l_c^3(x,y)}{h^3}\,,
\label{Q0}
\end{equation}
where $h$ is the side length of one voxel of the DTI dataset and $l_c$ a characteristic length estimated as 
\begin{equation*}
l_c(x,y)=\sqrt{\frac{tr( \mathbb D_W(x,y))h^2}{4l_1}}
\end{equation*} 
with $l_1$ the leading eigenvalue of the diffusion tensor $ \mathbb D_W$. In particular, for the side length $h$ we have $h=0.875$ mm, while $l_1$ is directly estimated from the DTI data in each voxel. For the ECs, we consider the situation of several blood vessels located in regions close to the tumor and where there is a high tissue density. From usual magnetic resonance imaging (MRI) and DTI scans, in fact, it is not possible to distinguish between vessels and ECM. Therefore, we assume the vasculature to be denser in the areas where more tissue is available. Defining
\begin{align}
&W_1(x,y)=2\,e^{-\frac{(x+2)^2}{0.5}}\,\sin^6{\left(\dfrac{\pi}{8}\,y\right)} \qquad \forall\, y\in[-5, 15]\,,\\[0.3cm]
&W_2(x,y)=2\,e^{-\frac{(x+12)^2}{0.5}}\,\sin^4{\left(\dfrac{\pi}{8}\,y\right)} \qquad \forall\, y\in[-35, -15]\,,\\[0.3cm]
&W_3(x,y)=2\,e^{-\frac{(x+5)^2+(y-15)^2}{0.75}}+2\,e^{-\frac{(x+5)^2+(y-20)^2}{0.75}}+2\,e^{-\frac{(x+10)^2+(y-20)^2}{0.75}}\,,
\end{align}
we set $W_0(x,y)=W_1(x,y)+W_2(x,y)+W_3(x,y)$. For the VEGF initial profile, we consider a Gaussian distribution, centered in the same point as tumor cells ${(x_{0,H},y_{0,H})=(-17,5)}$ and given by:
\[
H_0(x,y)=0.5\,e^{-\frac{(x-x_{0,H})^2+(y-y_{0,H})^2}{4}}\,.
\]
 Moreover, we assume that initially there is no necrotic matter, setting its initial value to zero everywhere in the domain. Figure \ref{In_Con_vegf} shows the initial conditions on the entire 2D brain slice, zooming then in the region $\bar{\Omega}=[-35, 5]\times[-15, 25]\subseteq \Omega$. \\
\begin{figure}[ht!]
\centering
\includegraphics[width=\textwidth]{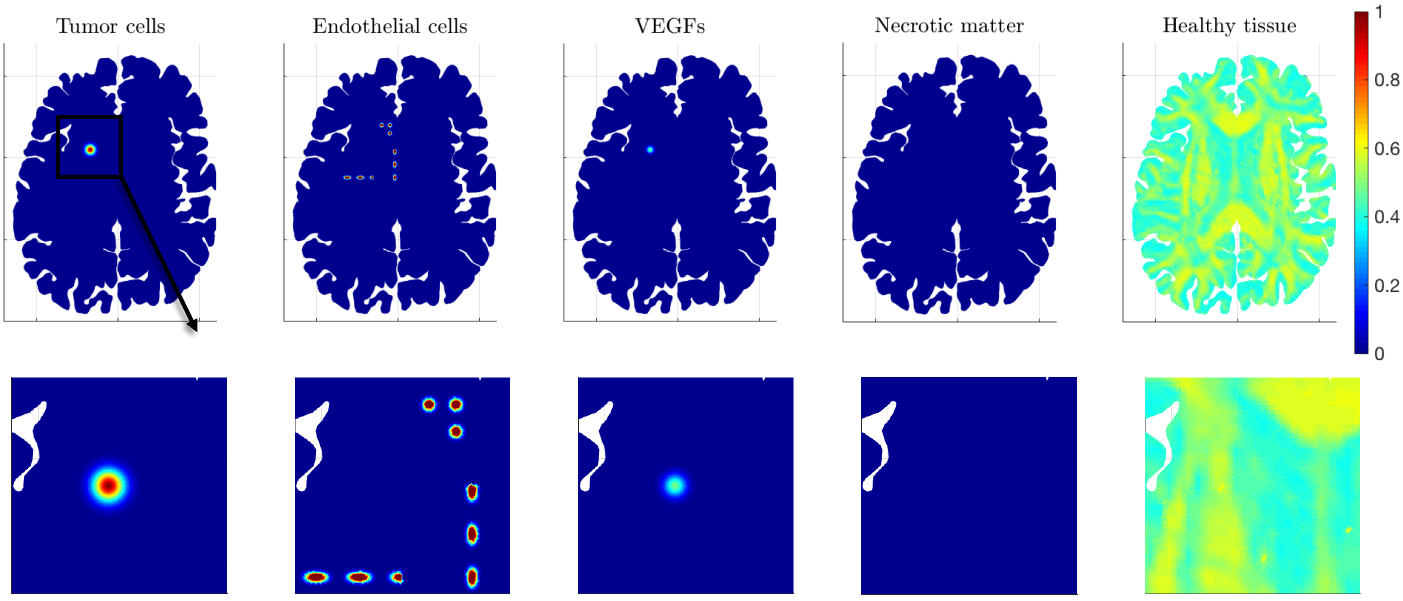}
\caption{\footnotesize{\textbf{Initial conditions of system \protect\eqref{mac_set_neu}.} The five columns refer to densities of glioma, ECs, VEGFs, necrotic matter, and healthy tissue, respectively. The initial densities in the bottom row are visualized on the zoomed domain $\bar{\Omega}=[-35, 5] \times [-15, 25]$.}}
\label{In_Con_vegf}
\end{figure} 

\noindent The numerical simulations are performed with a self-developed code in Matlab (MathWorks Inc., Natick, MA). The computational domain is a horizontal brain slice reconstructed from the processing of an MRI scan. The macroscopic tensor $\mathbb{D}_T(x)$ is pre-calculated using DTI data and the ODF for the fiber distribution function \eqref{qODF} (see Section \ref{CoefFun}). The DTI dataset was acquired at the Hospital Galdakao-Usansolo (Galdakao, Spain), and approved by its Ethics Committee: all the methods employed were in accordance with approved guidelines. We consider a Galerkin finite element scheme for the spatial discretization of the equations and an implicit Euler scheme for the time discretization. We run several simulations to test different scenarios, which can be summarized as follows. \\[0.1cm]
{\bf System evolution without any therapy}
\begin{itemize}
\item[{\bf (A)}] We study the dynamics of the five species involved in the macroscopic setting \eqref{mac_set_neu} in the absence of any  treatment. This means that the therapy doses ($d_r$ and $d_b$) and the chemotherapy killing rate ($k_c$) are set to zero and the tumor is let to grow and spread without any external impediment. 
\item[{\bf (B)}] We analyze the effect of the subcellular dynamics included in the EC equation, comparing the migration of these cells in three different cases. Precisely, we deduce for ECs two possible different equations: a simple diffusion model deduced from a kinetic equation where no subcellular dynamics are considered; a diffusion-advection model, where the transport term is obtained as in Ref.~\cite{conte_surulescu2020}. Then we compare the evolution of the ECs described in these two alternative ways with the approach proposed in these notes.
\end{itemize}
{\bf Effects of treatment}

The Stupp protocol\cite{stupp2005,stupp2009} has become standard of care for the treatment of gliomas. It consists of radiotherapy and concomitant chemotherapy with temozolomide. Radiotherapy (RT) is provided with the standard dose of 60 Gy delivered in 30 daily fractions of 2 Gy (Monday to Friday) over 6 weeks. RT planning is based on the definition of specific tumor volumes related to the tumor margins and identified by advanced imaging techniques. Chemotherapy is administered 7 days per week at 75 mg per square meter of body-surface area per day during radiotherapy, while different doses are used after the radiotherapy for six 28-day cycle. This second part of the chemotherapy treatment is also named adjuvant chemotherapy.  New treatments are also emerging to target molecules involved in various tumor mechanisms. Among them, angiogenesis inhibitors have been tested as anti-migratory agents for glioma. For instance, bevacizumab, a neutralizing monoclonal antibody against VEGFs, demonstrated a promising response in patients with recurrent malignant gliomas in combination with other drugs. However, the role and efficacy of bevacizumab are still debated\cite{reardon2011review}. 
\begin{itemize}
\item[{\bf (C)}] We study the effect of concurrent radio- and chemotherapy, which is the standard in clinical practice after tumor resection. The protocol we consider accounts for irradiation with 60 Gy in fractions of 2 Gy, simultaneously chemotherapy with temozolomide. The dosage of temozolomide (which is given orally) is according to the body surface of the patient: 75 mg/m$^2$ surface. The treatment schedule is sketched in Figure \ref{therapy_plan_scheme1}. With respect to the complete Stupp protocol, we neglect the adjuvant chemotherapy, i.e. the so-called "5/23 cycles", which consists of 5 days of temozolomide (150-200 mg/m$^2$) out of every 28 days and repeated for 6 times. Since our main aim is to compare the different treatment  schemes and to point out the effects of the anti-angiogenic therapy, the adjuvant chemotherapy would only be an additional (and perhaps confusing) factor. Moreover, in the real case reported in Section \ref{sec:real-data}, the example patient actually did not receive the adjuvant chemotherapy.

\begin{figure}[h!]
\centering
\includegraphics[width=0.9\textwidth]{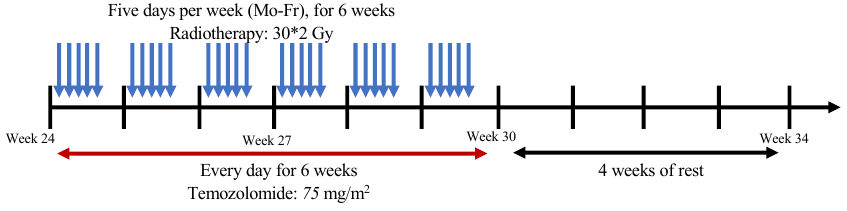}
\caption{\footnotesize{\textbf{Schematic representation of the therapeutic plan in experiment \textbf{(C)}.} The blue arrows refer to the radiotherapy administration, which is given in fractionated doses of $ d_r=2$ Gy per day; the red arrow identifies the period of chemotherapy administration, with a daily dose of $75$ mg $\cdot$ m$^{-2}$.}}
\label{therapy_plan_scheme1}
\end{figure} 

\item [{\bf (D)}] Following the protocol reported in the trials NCT01209442 and NCT01390948 (see the webpage \url{https://www.clinicaltrials.gov} for further details), we combined the treatment based on radio- and chemotherapy with an anti-angiogenic therapy with bevacizumab given every two weeks. The dosage of bevacizumab is accordingly to the body-weight of the patient: 10 mg/kg. The treatment schedule is sketched in Figure \ref{therapy_plan_scheme2}.
\begin{figure}[h!]
\centering
\includegraphics[width=0.9\textwidth]{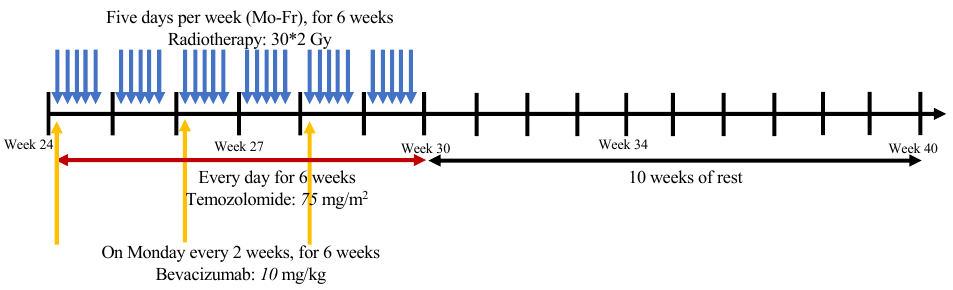}
\caption{\footnotesize{\textbf{Schematic representation of the therapeutic plan in experiment \textbf{(D)}.} The blue arrows refer to the radiotherapy administration, which is given in fractionated doses of $ d_r=2$ Gy per day; the yellow arrows refer to the anti-angiogenic therapy administration, given on a 2-weeks base at dose $d_b=10$ mg$ \cdot$ kg$^{-1}$;  the red arrow identifies the period of chemotherapy administration, with a daily dose of $75$ mg $\cdot$ m$^{-2}$.}}
\label{therapy_plan_scheme2}
\end{figure} 
\item [{\bf (E)}] Considering the trial AVF3708, whose results have been reported in Ref. ~\cite{friedman2009}, we test an adjuvant, second-line therapy with bevacizumab (10 mg/kg)  every 2 weeks on patients received prior radiotherapy and temozolomide. The treatment schedule for this experiment is sketched in Figure \ref{therapy_plan_scheme3}.
\begin{figure}[h!]
\centering
\includegraphics[width=0.9\textwidth]{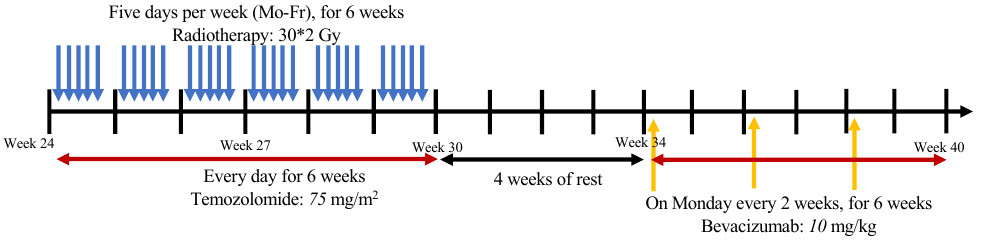}
\caption{\footnotesize{\textbf{Schematic representation of the therapeutic plan in experiment \textbf{(E)}.} The blue arrows refer to the radiotherapy administration, which is given in fractionated doses of $ d_r=2$ Gy per day; the yellow arrows refer to the adjuvant anti-angiogenic therapy administration, given on a 2-weeks base at dose $d_b=10$ mg$ \cdot$ kg$^{-1}$;  the red arrow identifies the period of chemotherapy administration, with a daily dose of $75$ mg $\cdot$ m$^{-2}$.}}
\label{therapy_plan_scheme3}
\end{figure} 
\end{itemize}
\subsection{System evolution with no therapy}
\label{NoTerSim}
\indent Starting with experiment {\bf(A)}, we simulate system \eqref{mac_set_neu} for the parameters listed in Table 1. To investigate the behavior of the different cell populations in the absence of therapy, we preliminary set $k_c=d_r=d_b=0$. This choice allows us to observe the tumor evolution over time without external intervention and to study the possible progression of the disease. The corresponding simulation results are shown in Figure \ref{NoTer_evolution}, where the five columns report the evolution of glioma density $M$, endothelial cell density $W$, VEGF concentration $H$, necrotic matter density $N_e$, and normal tissue density $Q$.
\begin{figure}[ht!]
\centering
\includegraphics[width=\textwidth]{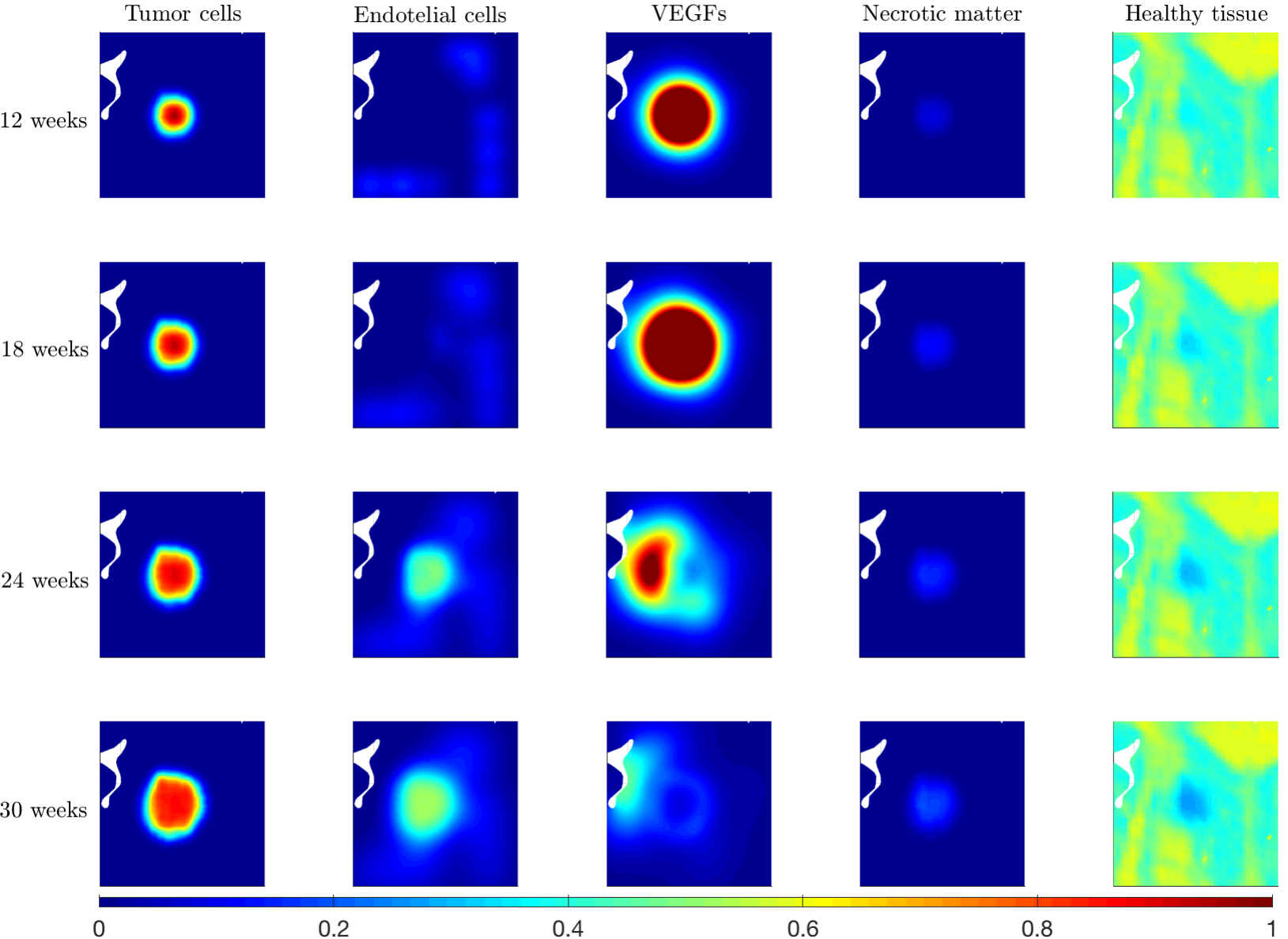}
\caption{\footnotesize{\textbf{Experiment (A).}} Numerical simulation of system \protect\eqref{mac_set_neu} with the parameters listed in Table 1 and without any treatment.}
\label{NoTer_evolution}
\end{figure} 
\noindent Tumor growth and spread, which mainly depend on the choice of the parameters $\lambda_0$ and $s$ (as also observed in Ref. ~\cite{conte_surulescu2020}), seems to be reasonably fast. Unlike in the articles\cite{conte_surulescu2020,Engwer2,Hunt2016}, here we are not relying on the go-or-grow hypothesis for the description of tumor evolution, whence we assume a phenotypically homogeneous tumor. In Figure \ref{NoTer_evolution}, we observe the growth and spread of the tumor mass in accordance with the distribution of healthy tissue (shown in the last column), which is crucially influencing the tumor tactic term, and of the vasculature. The EC dynamics also influence tumor proliferation through the function $g(W)$, which here is chosen to describe ECs as enhancing tumor proliferation and not directly controlling it (as considered, for instance, in Ref. ~\cite{conte_surulescu2020}). This choice is mainly related to the fact that tumor cells are still able to undergo mitosis (at a different rate) when there is a lack of nutrients; however, the presence of ECs substantially enhances cell proliferation. In Figure \ref{NoTer_evolution}, we also observe how ECs grow and diffuse, showing an evident migration toward the tumor population, due to the produced VEGFs. The latter is expressed by tumor cells and represents the chemotactic agent guiding ECs migration and, thus, mediating angiogenesis. In this case, since no therapy is included, the healthy tissue (last column of Figure \ref{NoTer_evolution}) is degraded only by glioma cells activity. More precisely, it is degraded by the acidity and matrix-degrading agents produced by tumor cells. However, here we wanted to avoid overloading the model by explicitly accounting for such variables (refer to Refs. ~\cite{conte_surulescu2020,kumar2020flux} for closely related problems with explicit acidity dynamics). The necrotic compartment collects this portion of degraded tissue. The tissue degradation shows an overall behavior comparable to the one observed in Ref.~\cite{conte_surulescu2020}.

\noindent To study the impact of including microscopic dynamics also in the EC equation, which led to the tactic term in \eqref{W0eq}, we compare the behavior of system \eqref{mac_set_neu} with that of two similar models where we change the EC description: one of them simply excludes the microscopic dynamics for ECs, deducing a macroscopic diffusion equation for these cells, while a second one considers EC migration as biased by the gradient of tumor cells (instead of the gradient of tumor-produced VEGF), similar to what is proposed in Ref. ~\cite{conte_surulescu2020}. In particular, in order to observe the impact on cell migration, we exclude EC proliferation from the three compared settings. In the first case, system \eqref{mac_set_neu} (without EC proliferation) is compared with an analogous system where the tumor equation \eqref{M0_eq} is coupled with the following reduced equation for ECs:
\begin{equation}
\partial_t W_1 -\nabla \cdot \left(\mathbb{D}_{EC}\nabla W_1\right)=0\,.
\label{Comp1}
\end{equation}
We refer to this system as $MC_1$ with $\mathbb{D}_{EC}$ given in \eqref{DEC}. In the second case, instead, system \eqref{mac_set_neu} (without EC proliferation) is compared with an analogous one, where the glioma equation \eqref{M0_eq} is coupled with the EC equation 
\begin{equation}
\partial_t W_2 -\nabla \cdot \left(\mathbb{D}_{EC}\nabla W_2\right)+\nabla \cdot\left(\chi_a(M_2)W_2\nabla M_2\right)=0\,.
\label{Comp2}
\end{equation}
We refer to this system as $MC_2$ and, here, $\chi_a(M_2):=\gamma_0a(M_2)\mathbb{D}_{EC}$, where $a(M_2)$ is the function accounting for the chemotactic effect of pro-angiogenic signals on ECs. Following Ref. ~\cite{conte_surulescu2020}, this function can be given as 
\[
a(M_2):=\chi_{a,0}\dfrac{K_{M_2}}{(K_{M_2}+M_2)^2}\,,
\]
with the parameter $\chi_{a,0}$ describing the duration between EC turnings damped by large glioma densities. This experiment {\bf(B)} is depicted in Figure \ref{Fig_Comp}, which collects the results of these comparisons. 
\begin{figure}[!htb]\hspace{-1.8cm}
    \centering
    \begin{minipage}{.5\textwidth}
        \centering
        \includegraphics[width=1.25\textwidth]{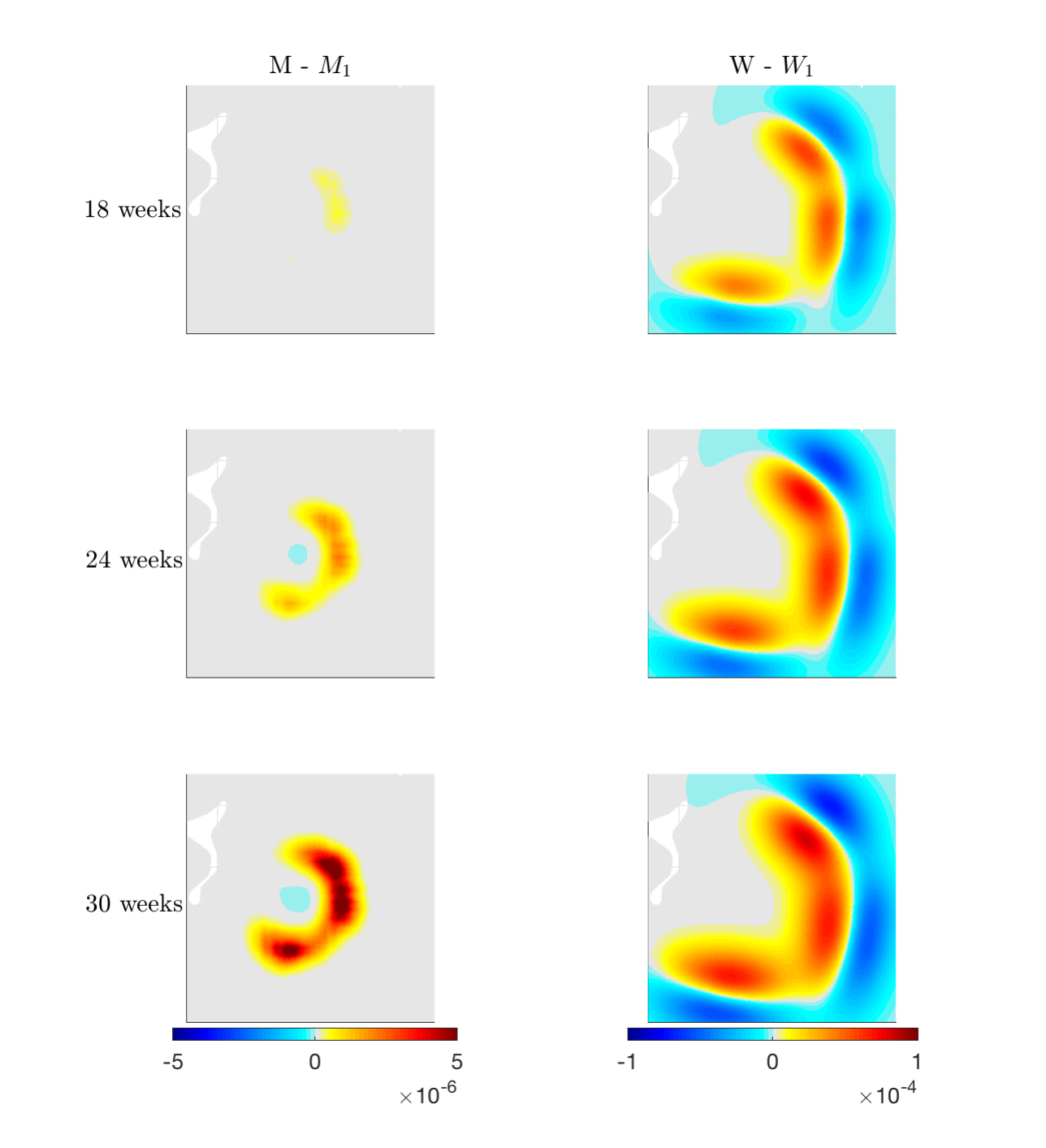}
    \end{minipage}%
    \hspace{0.8cm}\begin{minipage}{0.5\textwidth}
        \centering
       \includegraphics[width=1.25\textwidth]{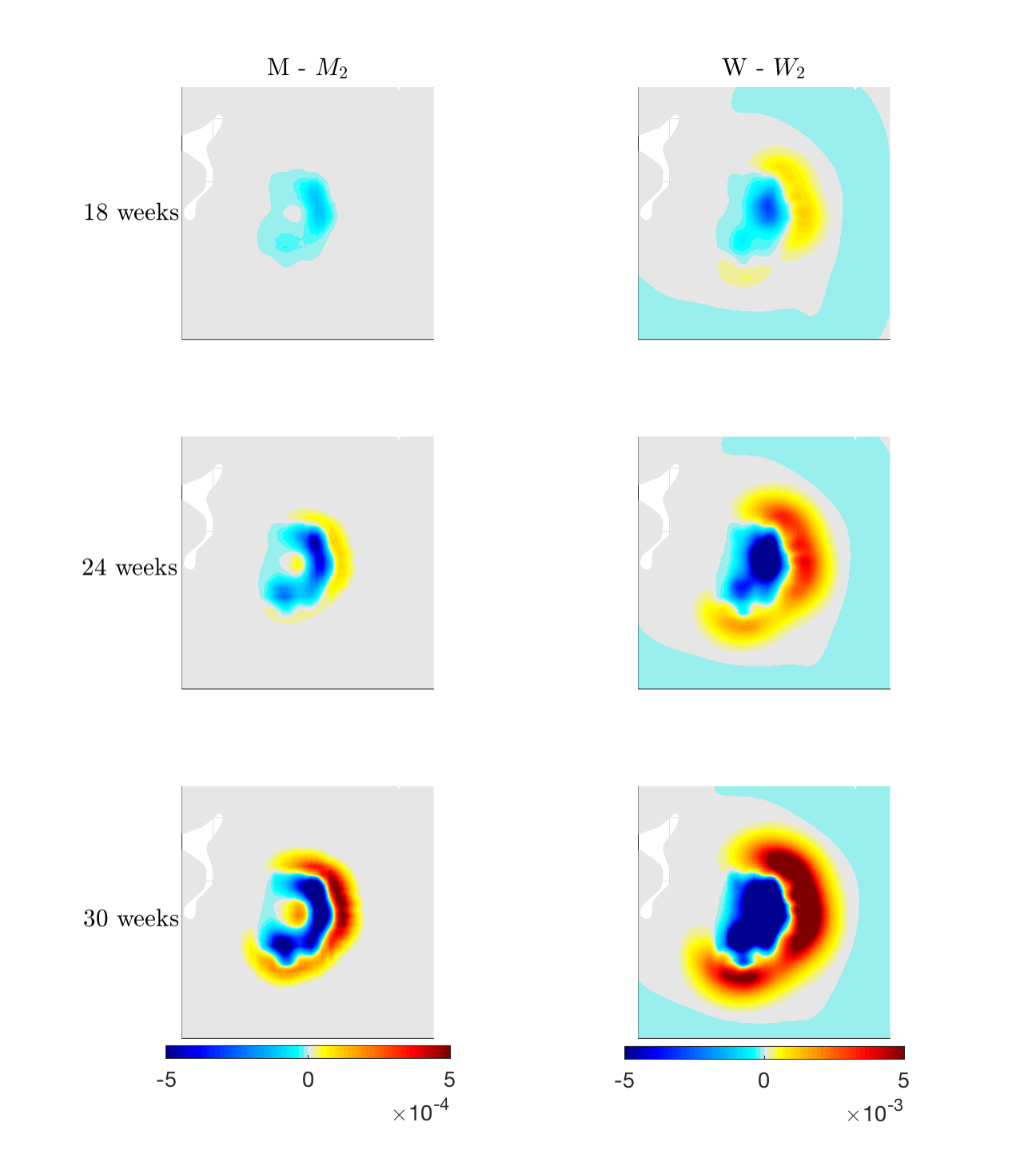}
    \end{minipage}
    \caption{\textbf{Experiment (B).} Differences between the respective solution components of Model \protect\eqref{mac_set_neu} and of the analogous one with equation \protect\eqref{Comp1} (first and second column) or equation \protect\eqref{Comp2} (third and fourth column) for ECs at 18 weeks (upper row), 24 weeks (middle row), and 30 weeks (lower row).}
    \label{Fig_Comp}
\end{figure}
Precisely, Figure \ref{Fig_Comp} shows the two different comparisons. The first two columns refer to the comparison between the model given by \eqref{mac_set_neu} and the one we named $MC_1$, while the third and fourth columns refer to the comparison between system \eqref{mac_set_neu} and model $MC_2$. We plot the differences between the solutions for the evolution of tumor ($M-M_i$) and endothelial cells ($W-W_i$) obtained with the different models. In the first comparison, when the equation for ECs is the simple diffusion equation \eqref{Comp1}, we observe - as expected - a stronger diffusive spread of ECs, which determines the negative values of the plotted difference (blue regions in the column $W-W_1$), while including the subcellular level for ECs leads to a more directed migration of these cells towards the tumor location, as shown by the positive values of the plotted difference (yellow/red regions in the column $W-W_1$). Correspondingly, as ECs enhance tumor proliferation, when their subcellular dynamics are included, we observe a higher density of tumor cells at the interface between the regions occupied by ECs and tumor cells (yellow/red regions in the column $M-M_1$). In the second comparison, we observe a much stronger directed migration of the ECs described by \eqref{Comp2}. In fact, higher values of EC density  (leading to negative values of the plotted difference in the column $W-W_2$) are shown mainly in the area affected by the tumor mass, whose gradient is driving the migration, when compared to the case in which EC motion is driven by the concentration of (tumor-produced) VEGFs. Once these growth factors are produced, they diffuse in the microenvironment, determining a lower tactic force on the endothelial cells. Tumor cell growth is, then, correspondingly affected, as shown in the column $M-M_2$. Precisely, there is enhanced  tumor proliferation in the core region in the case described by \eqref{Comp2}, while there is more proliferation around the external border in the case described by \eqref{mac_set_neu}.

\subsection{Effects of treatment}
\label{TerSim}
We study here the effects of the various therapeutic strategies described in the experiments \textbf{(C)}, \textbf{(D)} and \textbf{(E)} on the different system components. We denote with the index $S_1$ the solution components of experiment \textbf{(C)}, with $S_2$ the solution components of experiment \textbf{(D)}, and with $S_3$ the ones of experiment \textbf{(E)}. Starting from experiment {\bf (C)}, we first assess the impact of standard radio- and chemotherapy on the densities of glioma, ECs, necrotic matter, healthy tissue, and on the concentration of VEGFs. This numerical experiment is motivated by the fact that in the real case reported in Section \ref{sec:real-data} the patient was treated with this specific combined treatment. Figure \ref{therapy_plan_scheme1} shows a scheme of this treatment schedule.

\noindent With the same choice of parameters as that employed to generate  Figure \ref{NoTer_evolution} for experiment {\bf (A)}, we test the effect of the treatment plan depicted in Figure \ref{therapy_plan_scheme1}. The results of this experiment {\bf (C)} are shown in Figure \ref{Ter_evo_plan1}. The first row of Figure \ref{Ter_evo_plan1} (that also corresponds to the third row of Figure \ref{NoTer_evolution}) represents the state of the five species at the beginning of the treatment, while the second and third rows correspond to the situation after 3 and 6 weeks, respectively. Precisely, the third row represents the situation at the end of the combined treatment. In the last row, we show the system evolution after the resting period of 10 weeks, during which no therapy is applied.
	\begin{figure}[ht!]
\centering
\includegraphics[width=\textwidth]{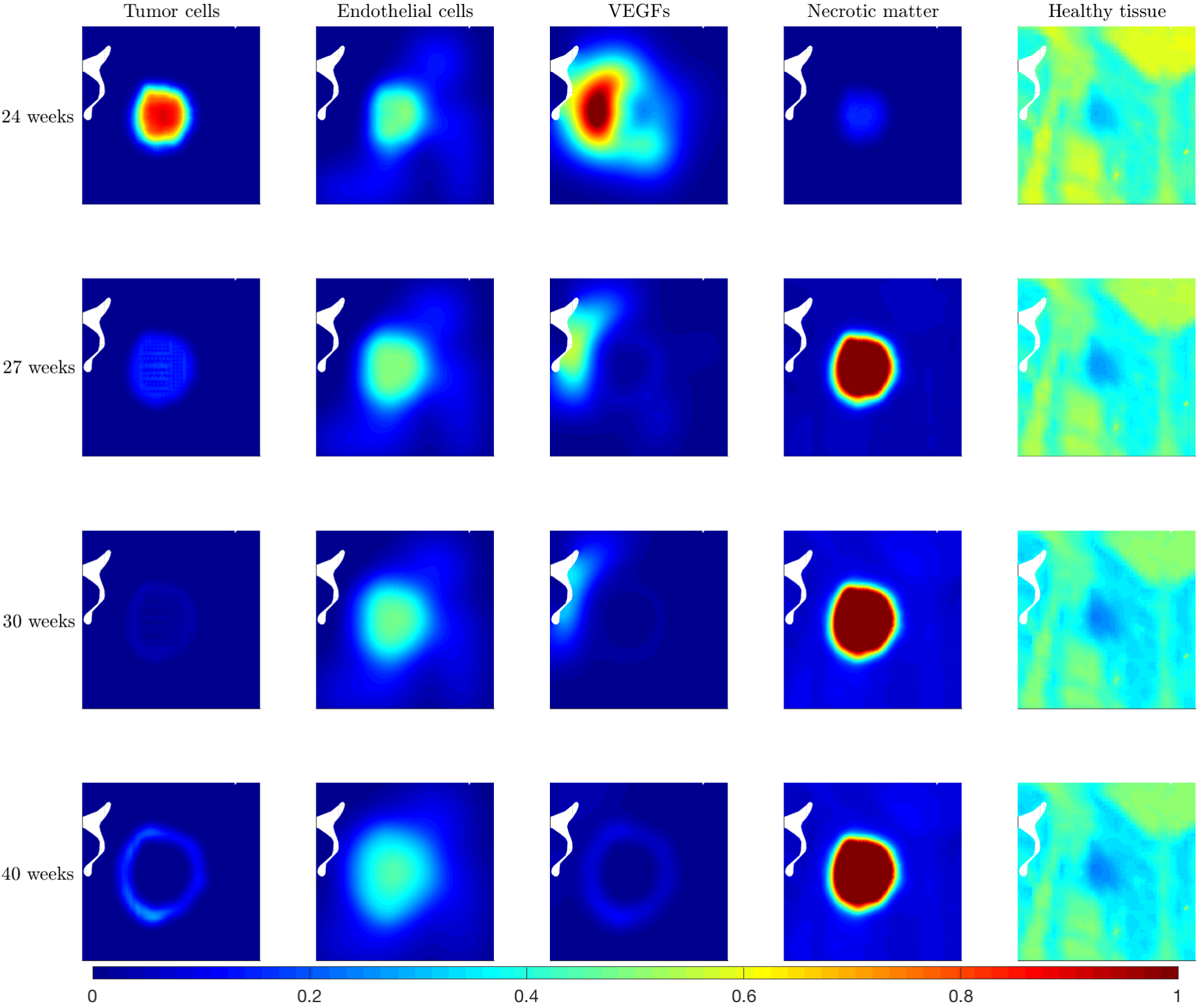}
\caption{\footnotesize{\textbf{Experiment (C).} Numerical simulation of system \protect\eqref{mac_set_neu} with the parameters listed in Table 1 and for the therapeutic plan schematized in Figure \ref{therapy_plan_scheme1}.}}
\label{Ter_evo_plan1}
\end{figure}
\noindent
Comparing Figures \ref{NoTer_evolution} and \ref{Ter_evo_plan1}, we immediately grasp the effects of the radio- and chemotherapy on the tumor population, whose density strongly decreases during treatment, while the density of the necrotic matter increases, as this component collects the effects of the therapy on tumor, ECs, and healthy tissue. The reduction in glioma density consequently affects VEGF production, with a decrease in its expression. The impact of the treatment is also evident in the evolution of the healthy tissue. In turn, the reduction of normal tissue affects the proliferation of cancer and endothelial cells. In fact, this depends on the availability of healthy tissue and, when the latter is excessively degraded, then proliferation is impaired. This effect can be observed by comparing the last row of Figures \ref{NoTer_evolution} and the third row of Figure \ref{Ter_evo_plan1}, which relates to the system evolution at 30 weeks (at the end of the treatment). 

\noindent In experiment \textbf{(D)} we investigate the overall effects of the treatment as given in experiment \textbf{(C)}, but supplemented with anti-angiogenic therapy. Precisely, the tumor is let to grow and spread for 24 weeks, allowing it to form a considerable neoplastic mass. Afterward, we apply a therapeutic plan consisting of a combination of radio- and chemotherapy, the latter involving temozolomide for inducing apoptosis of glioma cells, and anti-angiogenic therapy with bevacizumab for 6 weeks.  Temozolomide is administered orally every day at a standard constant dose of $75$ mg/m$^{2}$. Radiotherapy is applied 5 days per week (from Monday to Friday) for 6 weeks, and the total dose of $60$ Gy is fractionated into smaller doses of $2$ Gy per day. Anti-angiogenic therapy is administered at a dose of $10$ mg/kg intravenously every 2 weeks, thus, providing a total of 3 doses in the whole treatment period of 6 weeks. 
At the end of the treatment, we let the patient rest for 10  weeks and analyze how the tumor eventually reorganizes and evolves.  Figure \ref{therapy_plan_scheme2} shows the specific scheme of this treatment schedule. We also prolonged the simulations by 10 more therapy-free weeks, in order to better observe a possible tumor relapse. We consider the same choice of parameters as in Figure \ref{Ter_evo_plan1} for experiment {\bf (C)} to test the effect of the described combined therapy plan and we show the differences in the evolution of the solution components in Figure \ref{Ter_evo_plan12}. Precisely, we consider the differences between the species indicated with index $S_1$ (referring to the schedule in Figure \ref{therapy_plan_scheme1}) and those indicated with the index $S_2$ (referring to the schedule in Figure \ref{therapy_plan_scheme2}). Results are shown at 27 weeks (after three weeks of combined treatment), at 30 weeks (end of the treatment), at 40 weeks (after 10 weeks of no therapy), and finally at 50 weeks (after 10 more weeks without treatment).
\begin{figure}[ht!]
\centering
\includegraphics[width=\textwidth]{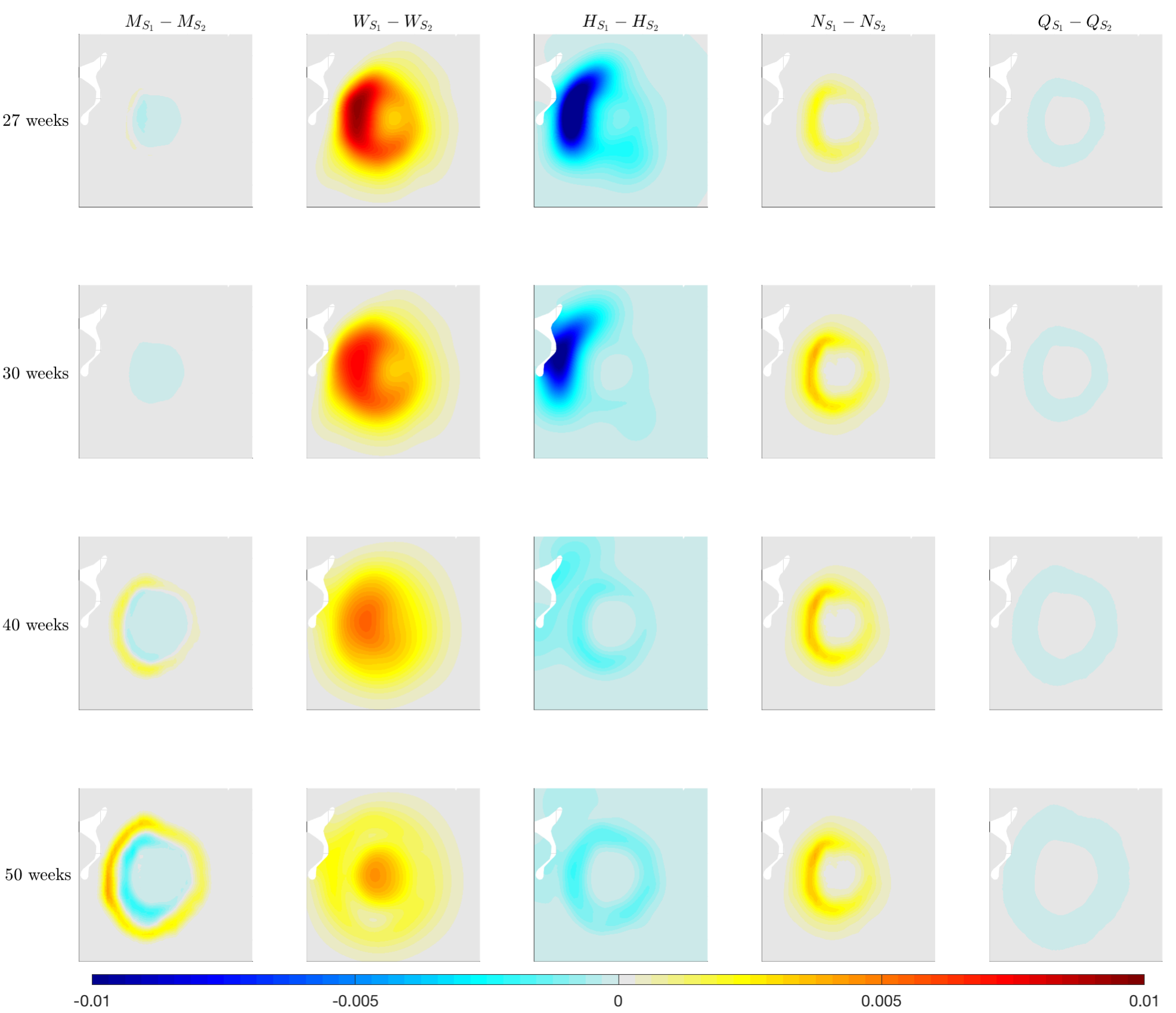}
\caption{\footnotesize{\textbf{Differences between experiment (C) and experiment (D).} Differences between the respective solution components of system \eqref{mac_set_neu} with the two different treatment schedules sketched in Figures \ref{therapy_plan_scheme1} and \ref{therapy_plan_scheme2} at 27 weeks (first row), 30 weeks (second row), 40 weeks  (third row), and 50 weeks (last row).}}
\label{Ter_evo_plan12}
\end{figure}
\noindent In Figure \ref{Ter_evo_plan12}, the effects of the additional anti-angiogenic therapy administered in the case $S_2$ can be observed. Firstly, the EC density is reduced when bevacizumab is administered, as shown by the positive values of the plotted difference (yellow/red regions in the column $W_{S_1}-W_{S_2}$). The consequent decrease in nutrient availability provided by the vasculature affects the tumor cell population and impacts the VEGF dynamics. Its expression is actually enhanced, thus leading to the negative values of the plotted difference (blue regions in the column $H_{S_1}-H_{S_2}$. The situation after 10 weeks of rest (third row of Figure \ref{Ter_evo_plan12}) suggests that in the long term the differences in the EC and VEGF densities reduce and, for the latter, they become almost negligible. Moreover, as a consequence of the decreasing EC density, when bevacizumab is administered the tumor mass is more confined, although featuring a slightly higher cell density at the tumor core (observed by the emergence of blue areas in those regions along with a yellow outer rim in the column $M_{S_1}-M_{S_2}$). The emergence of the latter induced us to simulate 10 more weeks of tumor evolution. The results are shown in the last row of Figure \ref{Ter_evo_plan12} and they confirm our guess that the difference between the $S_1$ and $S_2$ therapy plans becomes more visible after a longer intervention-free period. This is put in evidence by the clear enhancement of the yellow tumor rim: more glioma cells are spreading into the tumor-adjacent area if no anti-angiogenic treatment has been applied. Especially for glioma, this might be relevant, since very often the (infiltrative) margins of the treated tumor are the source of relapse. As with every therapy, one should also take into account the side effects, which should be balanced against reducing the tumor spread. The last column of the figure suggests that the effect of the anti-angiogenic drug on normal tissue is negligible. Concerning the necrotic matter, it correspondingly infers a substantial decrease when the treatment plan of Figure \ref{therapy_plan_scheme2} is administered, as observed in the column $N_{S_1}-N_{S_2}$. These simulations suggest that scheme $S_2$ might have some advantage against $S_1$, in line with some clinical results (refer e.g., to trial NCT01390948) showing a slight increase in the 1-year progression-free survival (from 67\% to 74\%) after the end of the treatment combining radio-, chemo-, and anti-angiogenic therapy. However, the clinical debate about the effects of bevacizumab is still open, as it did not seem to improve considerably the overall therapy outcome, despite its clear impact during the period of administration.

\noindent Finally, in experiment \textbf{(E)}, we analyze the effects of the treatment schedule sketched in Figure \ref{therapy_plan_scheme3}. Precisely, after letting the tumor grow for 24 weeks, we apply a combined treatment of radio- and chemotherapy. The former is applied 5 days per week (from Monday to Friday) for 6 weeks, at a fractionated dose of $2$ Gy per day (total dose of $60$ Gy), while the latter is based on temozolomide administration orally every day at a standard constant dose of $75$ mg/m$^{2}$. Then, after a resting period of 4 weeks, adjuvant anti-angiogenic therapy is applied at a standard dose of $10$ mg/kg intravenously every 2 weeks for other 6 weeks, thus providing a total of 3 doses during the whole treatment period.  With the same choice of parameters as in Figure \ref{Ter_evo_plan1} for experiment {\bf (C)}, we test the effects of the described combined therapy against the alternative therapy plan proposed in experiment {\bf (D)}. We show the differences in the evolution of the solution components for the two schedules in Figure \ref{Ter_evo_plan23}. Precisely, we consider the differences between the populations indicated with index $S_2$ (referring to the schedule in Figure \ref{therapy_plan_scheme2}) and the one indicated with the index $S_3$ (referring to the schedule in Figure \ref{therapy_plan_scheme3}). Results are shown at 27 weeks (after three weeks of radio- and chemotherapy), at 30 weeks (end of this treatment), at 40 weeks (after the resting period for the case $S_2$ or after the adjuvant anti-angiogenic treatment for the case $S_3$), and at 50 weeks, allowing for 10 more weeks without therapy for the follow-up.
\begin{figure}[ht!]
\centering
\includegraphics[width=\textwidth]{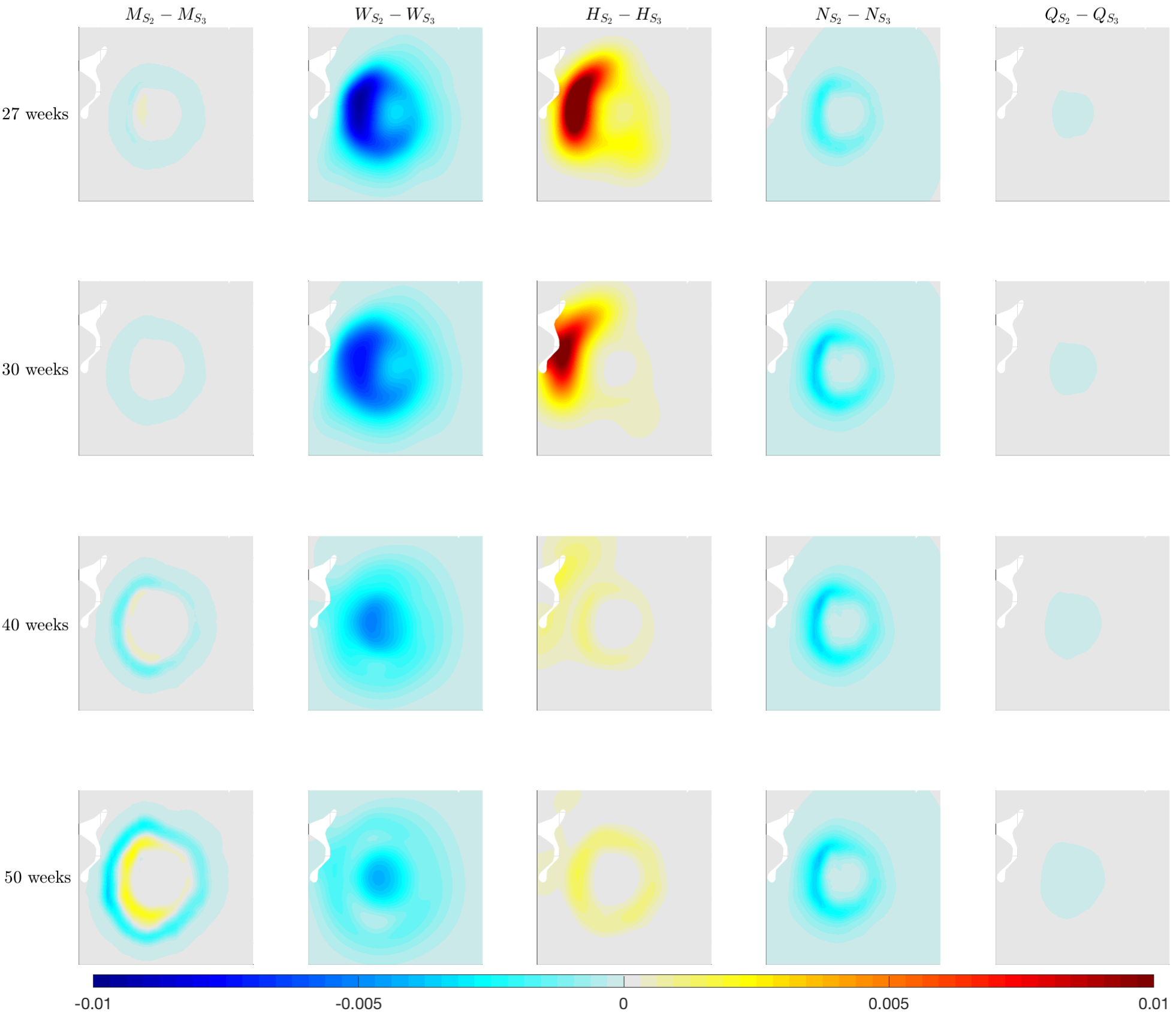}
\caption{\footnotesize{\textbf{Differences between experiment (D) and experiment (E).} 
Differences between the respective solution components of system \eqref{mac_set_neu} with the two different treatment schedules sketched in Figures \ref{therapy_plan_scheme2} and \ref{therapy_plan_scheme3} at 27 weeks (first row), 30 weeks (second row),  40 weeks (third row), and 50 weeks (last row).}}
\label{Ter_evo_plan23}
\end{figure}
\noindent From Figure \ref{Ter_evo_plan23} we can immediately notice the effects of the different schedules involving anti-angiogenic therapy on the EC population. Precisely, a concurrent administration of radio-, chemo-, and anti-angiogenic therapy seems to have a stronger effect on the EC density, which appears more reduced in case $S_2$ than in case $S_3$, as shown by the negative values of the plotted difference (blue regions in the column $W_{S_2}-W_{S_3}$). This has a clear impact on the tumor cell population, whose spread is reduced. Instead, the adjuvant anti-angiogenic therapy given after the resting period of 4 weeks seems to allow a stronger tumor infiltration into the surrounding areas, as shown by the blue outer ring of tumor cells in the column $M_{S_2}-M_{S_3}$. The reduction in EC density for the case $S_2$ correspondingly affects the VEGF component (yellow/red regions in the column $H_{S_2}-H_{S_3}$), which infers less uptake by ECs. The necrotic component sees a substantial increase in a much extended region when the treatment in Figure \ref{therapy_plan_scheme3} is administered, as tumor and EC activities are less controlled. The overall effect on the healthy tissue is, instead, minimal, in line with the results shown in Figure \ref{Ter_evo_plan12}.

\subsection{Application to real glioma patient data}\label{sec:real-data}
We perform further numerical simulations starting from real data of a glioblastoma patient in order to assess the predicted numerical outcome of the proposed therapeutic schemes. 

\noindent The patient aged 75 years first presented at the Department of Neurosurgery (Saarland University Medical Centre) with word-finding disorders, unsteady gait, and disorientation regarding time and situation. Computed tomography scans and subsequent magnetic resonance imaging revealed a left temporo-occipital brain lesion suspect for glioblastoma, which was confirmed pathologically after gross total resection. The histological work-up revealed glioblastoma cells, necrotic streaks, and prominent, partially hyperplastic endothelial cells. A combined radio- and chemotherapy was  recommended. Precisely, the treatment followed the standard protocol, applying 60 Gy in 30 fractions of 2 Gy over 6 weeks with simultaneous daily temozolomide (75 mg/m$^2$ surface area), analogous to the scheme shown in Figure \ref{therapy_plan_scheme1}. Due to severe thrombopenia, chemotherapy was stopped two days before the end of radiotherapy and was not resumed as adjuvant therapy.  

\noindent
We simulate a non-dimensionalized version of the macroscopic setting \eqref{mac_set_neu} on a horizontal brain slice reconstructed from the processing of the patient-specific MRI scans. The macroscopic tensor $\mathbb{D}_T(x)$ is pre-calculated using the patient DTI data and the bimodal von Mises-Fisher distribution\cite{PH13} for the fiber distribution function, whose expression is given by:
\[
q(x,\theta)=\dfrac{\delta}{2\pi}+(1-\delta)\dfrac{1}{4\pi I_0(k(x))}(e^{k(x)v_1\cdot \theta}+e^{-k(x)v_1\cdot \theta}).
\]
Here, the parameter $\delta\in[0,1]$ describes the inherent degree of randomized turning of the cell, while the function $k(x):=\kappa FA(\mathbb{D}_W(x))$, refers to the level of concentration around the dominant direction, with the parameter $\kappa$ quantifying the sensitivity of cells to the directional information given by the environment. $I_0(k(x))$ is the modified Bessel function of first kind and of order $0$ and $v_1$ is the leading eigenvector of $\mathbb{D}_W$.  In particular, the parameters $\delta$ and $\kappa$ are optimized using the patient data as in Ref. ~\cite{conte2020}, while the values for the other parameters are listed in Table 1. 

\noindent
We reconstruct the initial condition for tumor cells and necrotic matter from the provided patient segmentation data, where the different components of the tissue are classified into cerebrospinal fluid, gray matter, white matter, necrotic matter, edema, nonenhancing and enhancing tumor. In particular, for the tumor population we consider the areas occupied by both enhancing and  nonenhancing tumor. For the healthy tissue, we used the formulation given in \eqref{Q0}, with $l_c(x,y)$ and $h$ estimated from the patient DTI. For the initial density of ECs, which is neither explicitly available from the DTI brain data set nor the segmentation, we follow the choice proposed in Ref. ~\cite{dietrich2020}. Precisely, we choose the random variable
\[
W_0(x,y)=(10^{-3}+u_r)Q_0(x,y),
\]
where $u_r\sim U(0,10^{-3})$ denotes a random array derived from a continuous uniform distribution on $[0, 10^{-3}]$. Finally, for the VEGF initial profile, we assume a low concentration located in the region occupied by necrotic matter, where the lack of nutrients would favor VEGF production, i.e.,  
\[
H_0(x,y)=0.1\,N_{e,0}\,.
\]
Figure \ref{In_Con_RD} shows the initial conditions on the entire 2D brain slice.
\begin{figure}[ht!]
\centering
\includegraphics[width=\textwidth]{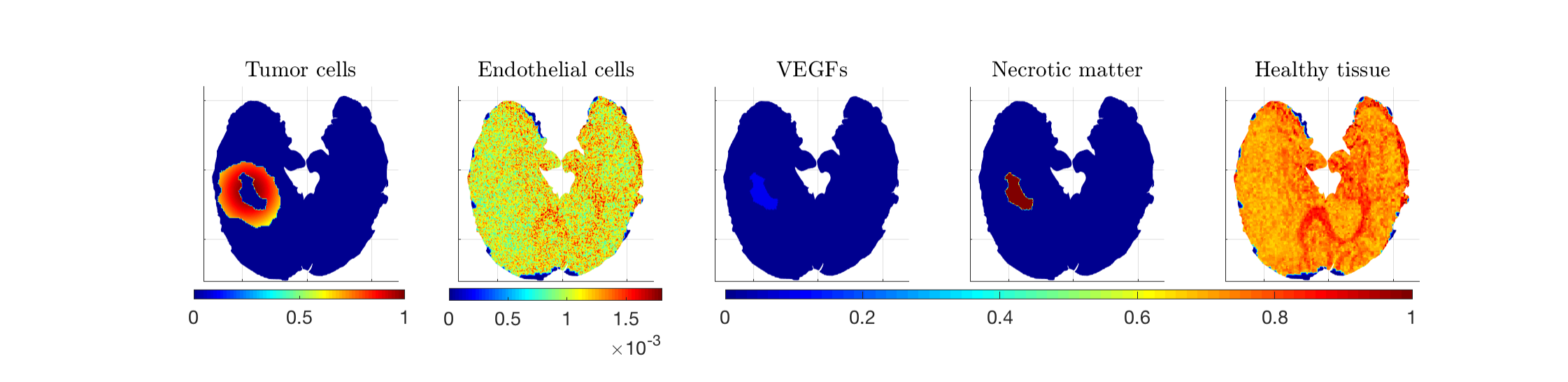}
\caption{\footnotesize{\textbf{Initial conditions of system \protect\eqref{mac_set_neu} reconstructed from real patient data.} The five columns refer to densities of glioma, ECs, VEGFs, necrotic matter, and healthy tissue, respectively.}}
\label{In_Con_RD}
\end{figure} 
\noindent We can notice that there are some errors or missed data in the border regions of the initial distributions derived from the DTI data (i.e., $Q_0$ and $W_0$). In fact, the main problem is that DTI measurements are usually recorded to analyze diffusion in white matter and, thus, data acquisition is terminated as soon as it reaches the margins of such regions. As a consequence, some values in these border areas can be missed and set to zero. However, the region of greater interest for our study, i.e., the one affected by the tumor mass, does not show particular problems to deal with. Concerning the treatment schedule, we consider the one described in Figure \ref{therapy_plan_scheme1}, i.e., a combination of chemo- and radiotherapy, as this was the plan actually administered to the patient. Precisely, we reconstructed from the data the dose distribution with which the patient was actually irradiated and we modified the terms \eqref{radio-term} modeling radiotherapy assuming a spatial-dependent radiotherapy dose $d_r(x,y)$. Figure \ref{Isodose} shows the isodose curves corresponding to radiation fractions of 10, 20, 30, 40, 50, and 60 Gy.  
\begin{figure}[h!]
\centering
\includegraphics[width=0.35\textwidth]{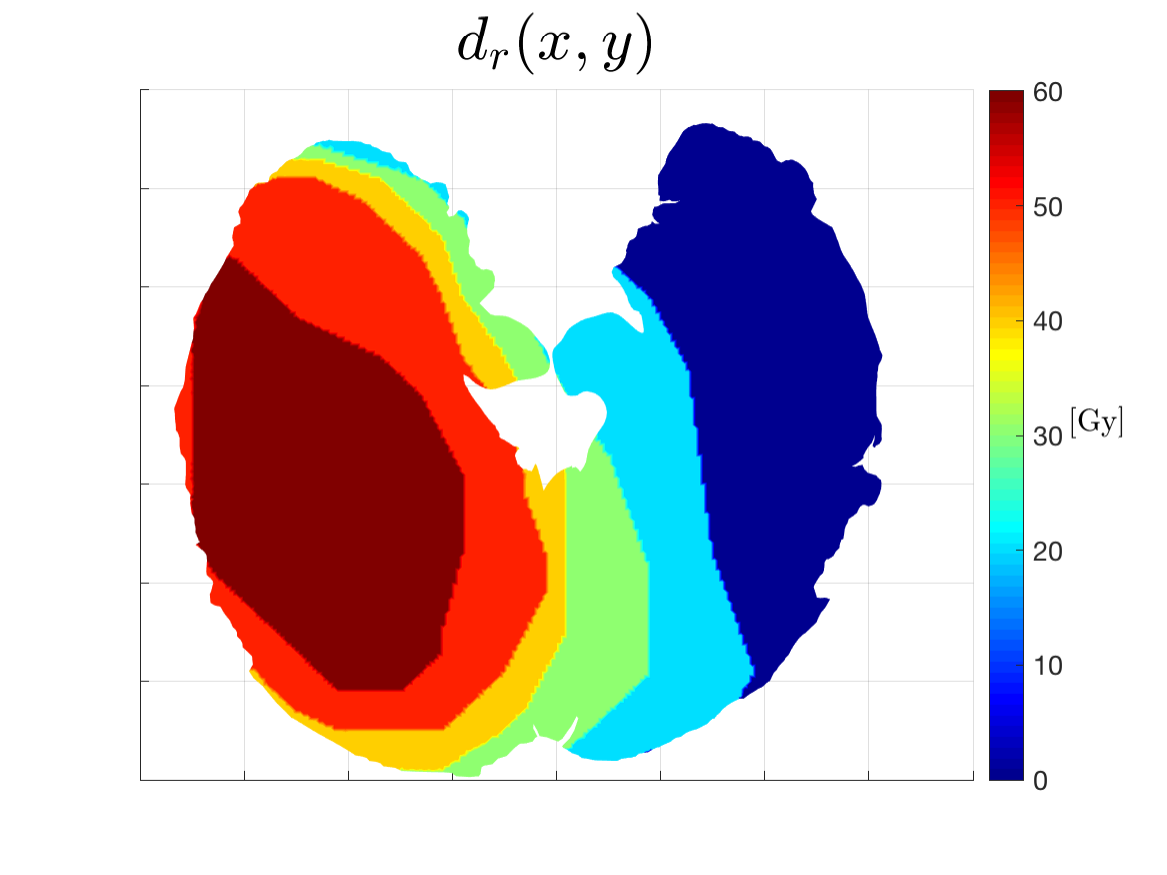}
\caption{\footnotesize{\textbf{Isodose curves of the radiotherapy treatment.} 
Level set curves referring to the specific dose distribution $d_r(x,y)$ with which the patient was irradiated.}}
\label{Isodose}
\end{figure}
\noindent Starting from the initial conditions given in Figure \ref{In_Con_RD}, we simulate the system evolution for 10 weeks (6 weeks of combined therapy + 4 weeks of rest). Results are given in Figure \ref{EvoRD}.
\begin{figure}[h!]
\centering
\includegraphics[width=\textwidth]{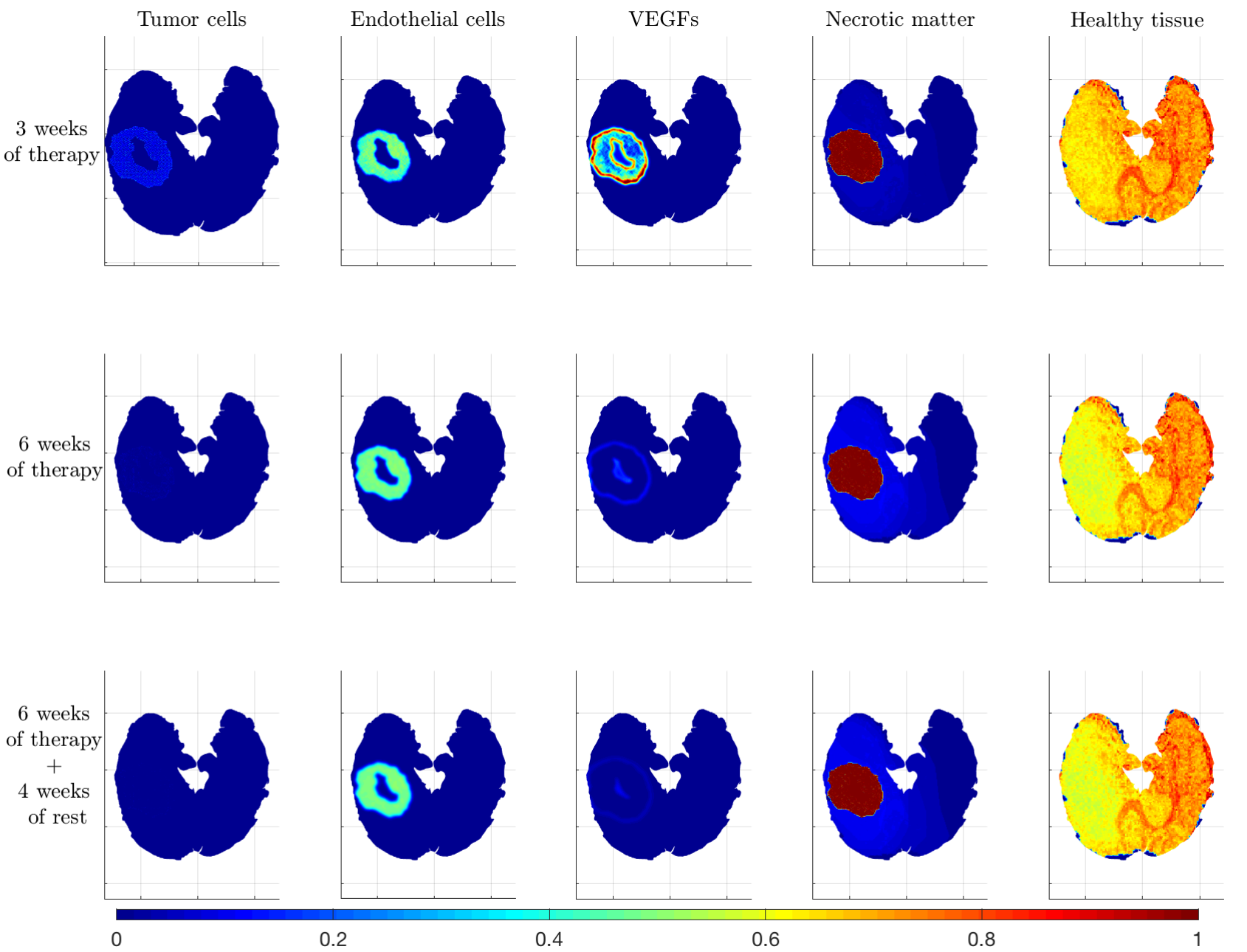}
\caption{\footnotesize{\textbf{Experiment (C) on real patient data.} 
Numerical simulation of system \protect\eqref{mac_set_neu} with the parameters listed in Table 1, for the therapeutic plan schematized in Figure \ref{therapy_plan_scheme1}, and on the basis of the provided patient data.}}
\label{EvoRD}
\end{figure}
\noindent Firstly, we observe the tumor dynamics (first column of Figure \ref{EvoRD}), which is characterized by a strong reduction of the tumor density in the area originally occupied by the tumor mass. This shows the efficacy of the combined treatment in reducing the overall tumor volume. Moreover, we notice that, even after the resting period of 4 weeks, no evident regrowth of the tumor mass emerges. Concerning VEGFs and ECs, we observe that the production of VEGFs by tumor cells and their accumulation in the region originally affected by the tumor determines a correspondingly increasing concentration of ECs in the same area. In turn, the increasing density of ECs, together with the reduction of tumor density, leads to a noticeable decrease in VEGF concentration, as shown in the third column of Figure \ref{EvoRD}. In particular, the EC density (especially after three weeks of therapy) shows a certain level of heterogeneity, which may be due to the random distribution we chose as initial condition for this population and which determines a corresponding heterogeneity in the VEGFs concentration (first plot in the third column of Figure \ref{EvoRD}). An outer rim of VEGFs seems to remain after the 10 weeks and it may be due to a corresponding low-dense outer rim of tumor cells surviving the treatment. Looking at the evolution of necrotic matter and healthy tissue, we notice the effects of having assumed a space-dependent radiotherapy dose distribution. In fact, the degradation of the healthy tissue is not homogenous and the further the region from the tumor area, the smaller the effect of radiations on healthy tissue. In fact, the most distant regions (e.g.  those in the right hemisphere) do not seem to be affected by the therapy. The effects of both heterogeneous tissue degradation and tumor cell death are, then, collected in the necrotic compartment.
\section{Discussion}\label{conclusion}
The model proposed in this note is a therapy-oriented development of that considered in Ref. ~\cite{conte_surulescu2020} and aligns to other approaches\cite{conte2020,CEKNSSW,Corbin2018,Engwer2,EHS,Engwer,Hunt2016}. Unlike Refs. ~\cite{conte_surulescu2020,Engwer} we do not differentiate here between moving and proliferating cancer cells, but account for a single population of cells forming the tumor, interacting with a population of endothelial cells representing the vascularization, along with a macroscopic description of brain tissue degradation and the dynamics of VEGF and necrotic matter. In fact, the study in  Ref. ~\cite{conte_surulescu2020} showed (at least in our framework) that the main effect of splitting the tumor population into moving and resting (and proliferating) cells was the slower evolution of tumor mass, but qualitatively it behaved similarly to the situation with one population of moving and proliferating cancer cells. Therefore, in this note we do not differentiate between the two cell phenotypes and focus instead on therapy effects. As mentioned, the main physical components of the tumor environment are the anisotropic brain tissue as well as brain-own and tumor-generated vasculature. Explicitly modeling the necrotic component of the neoplasm is motivated by its relevance for assessing the tumor stage and for the segmentation needed for treatment planning. The model investigates the impact of including microscopic dynamics also in the equation for endothelial cells, which has not been done before, comparing our approach with other two possible descriptions of EC dynamics. Then, we analyze via simulations various therapy approaches and their effects on tumor development and on normal tissue, thereby following treatment schedules addressed in clinical studies available in the literature. Although this has been done in Ref. ~\cite{Hunt2016} in a rather rudimentary way, we do not include tumor resection as part of the therapy; the problems have already been stated in that reference: it is hardly possible to provide a proper mathematical characterization of the tissue and neoplasm dynamics in the resected region in a continuous manner and the solution infers sharp discontinuities, which are difficult to handle. Considering the real data employed in Subsection \ref{sec:real-data}, the simulations showed qualitatively reasonable results, however, a quantitative assessment could not be performed, due to missing data, as that patient's therapy was stopped.

\noindent
The macroscopic system \eqref{mac_set_neu} obtained by parabolic upscaling carries therapy-related coefficients in the taxis as well in the reaction terms of glioma and EC dynamics. Obtaining such explicit dependencies would not be possible if a direct, phenomenological modeling on the macroscopic level were performed. Indeed, unlike previous works involving glioma therapy in their models (e.g. see\cite{Holdsworth_2012,Rockne2010}), our approach starts by describing treatment effects on the microscopic level of single cells interacting with their environment through transmembrane units, which can be impaired by some chemotherapeutic (e.g., anti-angiogenic) drug or reduced by the restrained availability of insoluble ligands as a consequence of ionizing radiation. As the various therapies are in fact acting on this microlevel, a multiscale modeling approach within the KTAP framework seems to be particularly adequate for a more careful description of the way they affect the targeted cell populations and of the respective response of tissues, as observed in Ref. ~\cite{Hunt2016}. Notwithstanding the issues related to parameter estimation and model calibration, such bottom-up approach enables including a great level of detail, however without the costly simulations typically related to discrete modeling efforts.

\noindent
System \eqref{mac_set_neu} with its initial and boundary conditions is also interesting from a mathematical viewpoint: it describes a novel sort of taxis cascade, in which tumor cells are chemotactically following ECs, which in turn bias their motion towards gradients of a chemical signal (VEGF) produced by tumor cells and depleted by themselves. The global well-posedness of a system featuring this complexity has not been investigated and raises manifold challenges, due to the intricate couplings and nonlinearities involved in the source and motility terms. The passage from KTEs to the macroscopic dynamics has been done here in a merely formal way, the rigorous convergence is still open and might probably use some of the ideas in   
Ref. ~\cite{ZS22}, where a much simpler system has been considered.

\noindent
In Ref. ~\cite{knobe2021} we commented about the feasibility of using such multiscale models to predict tumor spread and establish CTV and PTV margins for treatment planning. Those observations still apply here - the main issue remains the relatively large number of parameters and therewith related uncertainties. Although a proper calibration of the many involved parameters is required to support the quantitative validity of the proposed setting, we used a real patient data set to show that this model is able to provide qualitatively reasonable results. Indeed, the qualitative behavior observed with the last numerical test appears to be in good agreement with what the medical doctors observe in reality. Moreover, the increasingly fast development of biomedical imaging, computing power, and technology for necessary cell biology experiments will provide a means to assess at least some of the missing quantitative information. On the other hand, such multiscale models seem to offer an adequate frame for studying the effects of various therapy ansatzes. In fact, chemo- and radiotherapy primarily act on the level of single cells and ultimately lead to the observed effects on the whole tumor and this mathematical approach allows us to account for dynamics on both levels in a reasonably detailed manner. Our numerical experiments also suggest that letting the clinical studies have a longer follow-up might provide useful information about the tumor behavior after ceasing the actual therapy. That would presumably lead to higher costs, but these might be justified by the achieved understanding. Mathematical models could help in identifying the adequate duration of such studies. Moreover, they have the potential to investigate a great variety of therapeutic scenarios (of which we showed here just a few examples) in an unprecedented complexity and accuracy - provided the necessary quantitative information becomes available. Intra- and interdisciplinary studies including such models are called upon to shed light on the intricate biological processes associated with tumor growth, expansion, and treatment response.

\appendix
\section{Non-dimensionalization}
\label{adim_sis}
To proceed with the non-dimensionalization, we firstly observe that the variables $M$, $W$, $N_e$ involved in systems (\ref{mac_set_neu}) are expressed in cells/mm$^3$, $Q$  in mg/mm$^3$, while the concentration of VEGF $H$ is given in mol/liter (=:M). The reference values we use for the adimensionalization are listed in Table 2. In particular, we rescale the tumor, EC, and necrotic matter densities with respect to their carrying capacities, i.e., $M_{c,0}=K_M$, $W_{c,0}=K_W$, and $N_{e,0}=K_{Ne}$.

\begin{table} [!h] \label{ref_val}
\begin{center}
\caption{Reference variables}
{\begin{tabular}{@{}c|c|c|c@{}}\toprule
Parameter & Description & Value (units) & Source \\
\midrule
   T & time & $1$ (d) & \\
   L & length & $0.875$ (mm)& \\
$M_{c,0}$  & tumor cell density & $10^{5}$ (cell$\cdot$mm$^{-3}$)   & \cite{conte_surulescu2020}  \\
$W_{c,0}$ & ECs density &$10^{5}$ (cell$\cdot$mm$^{-3}$)&\cite{conte_surulescu2020}  \\
$N_{e,0}$ & Necrotic matter density &$10^{5}$ (cell$\cdot$mm$^{-3}$)&\cite{conte_surulescu2020}  \\
$H_{c,0}$  & VEGF concentration & $10^{-9}$ (M)& \cite{Takano} \\
$Q^*$  & healthy tissue density & $10^{-3}$ (mg$\cdot$mm$^{-3}$) &\cite{conte_surulescu2020}  \\
    \bottomrule
    \end{tabular}}
    \end{center}
\end{table}

\noindent
We nondimensionalize the partial differential equations introduced above as follows:
\begin{equation*}
\tilde{t}=\dfrac{t}{T},\quad \tilde{x}=\dfrac{x}{L},\quad\tilde{M}=\dfrac{M}{M_{c,0}},\quad\tilde{W}=\dfrac{W}{W_{c,0}},\quad\tilde{N_e}=\dfrac{N_e}{N_{e,0}},\quad\tilde{H}=\dfrac{H}{H_{c,0}},\quad\tilde{Q}=\dfrac{Q}{Q^*}\,.
\end{equation*}
A proper scaling of the parameters involved in the macroscopic setting leads to:
\begin{equation*}
\begin{split}
&\tilde{k}_W^+=\dfrac{k_W^+}{\lambda_0}, \quad \quad \tilde{k}_Q^+=\dfrac{k_Q^+}{\lambda_0},\quad\quad\tilde{k}_1^-=\dfrac{k_1^-}{\lambda_0},\quad\quad\tilde{k}_H^+=\dfrac{\bar{k} _H^+}{\gamma_0}, \quad\quad\tilde{\mu}_{M,0}=\mu_{M,0}\,T,\\[0.3cm]
&\tilde{\lambda_1}=\dfrac{\lambda_1}{k_1^-},\quad\quad \tilde{k}_H^-=\dfrac{\bar{k}_H^-}{\gamma_0},\quad\quad \tilde{\gamma}_1=\dfrac{\gamma_1}{\gamma_0},\quad\quad \tilde{\mu}_{W,0}=\bar{\mu}_{W,0}\,T, \quad \quad \tilde{c}_{2}=c_{2}\,T,\\[0.3cm]
&\tilde{\mathbb{D}}_T=\dfrac{1}{\lambda_0}\dfrac{T}{L^2}\mathbb{D}_T,\quad\quad\tilde{\mathbb{D}}_{EC}(x)=\dfrac{T}{L^2}\,\mathbb{D}_{EC}(x),\quad\quad\tilde{D}_H=\dfrac{T}{L^2}D_H,\quad\quad \tilde{c}_{1}=\dfrac{c_{1}}{H_{c,0}}T.
\end{split}
\end{equation*}
\noindent
If we drop the tilde (" $\tilde{ }$ ") in the new variables and parameters, then the differential equations in sistem (\ref{mac_set_neu}) keep the same form, with the properly rescaled expressions for following functions:
\begin{equation*}
\begin{split}
&\tilde{\mu}(\tilde{M},\tilde{W},\tilde{N}_e):=\tilde{\mu}_{M,0}\,\left(1-\tilde{M}-\tilde{N}_e\right)g(\tilde{W})\,,\qquad \tilde{\mu}_{W,0}(d_b)=\tilde{\mu}_{W,0}\left(1+\dfrac{1}{1+d_b^2}\right)\,,\\[0.3cm]
&\tilde{\mu}_W(\tilde{W},\tilde{Q},\tilde{d_b})=\tilde{\mu}_{W,0}(d_b)(1-\tilde{W})\tilde{Q}\,,\qquad  \tilde{B}_p(\tilde{W},\tilde{Q})=\left(\tilde{k}_W^+S_W\tilde{W}+\tilde{k}_Q^+S_Q\tilde{Q}+\tilde{k}_1^-\right)\,,\\[0.3cm]
&\tilde{B}_w(\tilde{H})=(\tilde{k}_H^+\tilde{H}+\tilde{k}_H^-).\\
\end{split}
\end{equation*}

\section*{Acknowledgment}
MC acknowledges funding by the  by Ministry of Education, Universities and Research, through the MIUR grant Dipartimento di Eccellenza 2018-2022, Project no. E11G18000350001. This work was also supported by National Group of Mathematical Physics” (GNFM-INdAM). MC acknowledges also funding by the Basque Government through the BERC 2018- 2021 program, by the Spanish State Research Agency through BCAM Severo Ochoa excellence accreditation SEV-2017-0718 and by the MINECO- Feder (Spain) through the research grant number RTI2018-098850-B-I00. MC has received funding from the European Union’s Horizon 2020 research and innovation programme under the Marie Sk\l{}odowska-Curie grant agreement No. 713673. The project that gave rise to these results received the support of a fellowship from “a Caixa” Foundation (ID 100010434). The fellowship code is LCF/BQ/IN17/11620056. This work has been partially supported by the State Research Agency of the Spanish Ministry of Science and FEDER-EU, project PID2022-137228OB-I00 (MICIU/AEI /10.13039/501100011033); by Modeling Nature Research Unit, Grant QUAL21-011 funded by Consejería de Universidad, Investigaci\'on e Innovaci\'on (Junta de Andalucía) (MC). CS was partially supported by the Federal Ministry of Education and Research BMBF, project \textit{GlioMaTh} 05M2016.

\newcommand{\noopsort}[1]{}
\addcontentsline{toc}{section}{References}
\bibliographystyle{plain}
\bibliography{bibl2}

@Article{EHS,
	author	= {C. Engwer and A. Hunt and C. Surulescu},
	title   = {Effective equations for anisotropic glioma
	spread with proliferation: a multiscale approach},
	journal = {IMA Journal of Mathematical Medicine and Biology},
	volume = {33},
	year	= {2016},
	pages= {435--459}
}

@article{burini2019multiscale,
  title={A multiscale view of nonlinear diffusion in biology: From cells to tissues},
  author={Burini, Diletta and Chouhad, Nadia},
  journal={Mathematical Models and Methods in Applied Sciences},
  volume={29},
  number={04},
  pages={791--823},
  year={2019},
  publisher={World Scientific}
}

@article{bellomo2021life,
  title={What is life? A perspective of the mathematical kinetic theory of active particles},
  author={Bellomo, N. and Burini, D. and Dosi, G. and Gibelli, L. and Knopoff, D. and Outada, N. and Terna, P. and Virgillito, M.E.},
  journal={Mathematical Models and Methods in Applied Sciences},
  volume={31},
  number={09},
  pages={1821--1866},
  year={2021},
  publisher={World Scientific}
}

@article{bellomo2022towards,
  title={Towards a mathematical theory of behavioral human crowds},
  author={Bellomo, N. and Gibelli, L. and Quaini, A. and Reali, A.},
  journal={Mathematical Models and Methods in Applied Sciences},
  volume={32},
  number={02},
  pages={321--358},
  year={2022},
  publisher={World Scientific}
}

@article{albi2019vehicular,
  title={Vehicular traffic, crowds, and swarms: From kinetic theory and multiscale methods to applications and research perspectives},
  author={Albi, G. and Bellomo, N. and Fermo, L. and Ha, S.-Y. and Kim, J. and Pareschi, L. and Poyato, D. and Soler, J.},
  journal={Mathematical Models and Methods in Applied Sciences},
  volume={29},
  number={10},
  pages={1901--2005},
  year={2019},
  publisher={World Scientific}
}

@techreport{prigogine1971kinetic,
  title={Kinetic theory of vehicular traffic},
  author={Prigogine, I. and Herman, R.},
  year={1971}
}

@article{klar1997enskog,
  title={Enskog-like kinetic models for vehicular traffic},
  author={Klar, A. and Wegener, R.},
  journal={Journal of statistical Physics},
  volume={87},
  number={1},
  pages={91--114},
  year={1997},
  publisher={Springer}
}

@Article{PH13,
	author	= {K. Painter and T. Hillen},
	title   = {Mathematical modelling of glioma growth: the use of diffusion tensor imaging ({DTI}) data to predict the anisotropic pathways of cancer invasion},
	journal = {Journal of Theoretical Biology},
	volume = {323},
	year	= {2013},
	pages= {25--39}    
}

@article{Giese2003,
	doi = {10.1200/jco.2003.05.063},
	url = {https://doi.org/10.1200/jco.2003.05.063},
	year = {2003},
	publisher = {American Society of Clinical Oncology ({ASCO})},
	volume = {21},
	number = {8},
	pages = {1624--1636},
	author = {A. Giese and R. Bjerkvig and M.E. Berens and M. Westphal},
	title = {Cost of Migration: Invasion of Malignant Gliomas and Implications for Treatment},
	journal = {Journal of Clinical Oncology}
}

@article{giese-etal96,
	title={Migration of human glioma cells on myelin.},
	author={A. Giese and L. Kluwe and H. Meissner and E. Michael and M. Westphal},
	journal={Neurosurgery},
	volume={38},
	number={},
	pages={755--764},
	year={1996},
}

@article{Engwer4,
	title={On a structured multiscale model for acid-mediated tumor invasion: The effects of adhesion and proliferation.},
	author={C. Engwer and C. Stinner and C. Surulescu},
	journal={Mathematical Models and Methods in Applied Sciences},
	volume={27},
	year={2017}, 
	pages={1355-1390}
}

@article{Engwer2,
	doi = {10.1007/s00285-014-0822-7},
	url = {https://doi.org/10.1007/s00285-014-0822-7},
	year = {2014},
	publisher = {Springer Science and Business Media {LLC}},
	volume = {71},
	number = {3},
	pages = {551--582},
	author = {C. Engwer and T. Hillen and M. Knappitsch and C. Surulescu},
	title = {Glioma follow white matter tracts: a multiscale {DTI}-based model},
	journal = {Journal of Mathematical Biology}
}

@article{Engwer,
	doi = {10.3934/mbe.2015011},
	url = {https://doi.org/10.3934/mbe.2015011},
	year = {2016},
	publisher = {American Institute of Mathematical Sciences ({AIMS})},
	volume = {13},
	number = {2},
	pages = {443--460},
	author = {C. Engwer and M. Knappitsch and C. Surulescu},
	title = {A multiscale model for glioma spread including cell-tissue interactions and proliferation},
	journal = {Mathematical Biosciences and Engineering},
}

@article{Berens1999,
	doi = {10.1038/sj.neo.7900034},
	url = {https://doi.org/10.1038/sj.neo.7900034},
	year = {1999},
	publisher = {Elsevier {BV}},
	volume = {1},
	number = {3},
	pages = {208--219},
	author = {M.E. Berens and A. Giese},
	title = {{\textquotedblleft}...those left behind.{\textquotedblright} {B}iology and Oncology of Invasive Glioma Cells},
	journal = {Neoplasia}
}

@article{Plaza,
	doi = {10.1007/s00285-018-1323-x},
	url = {https://doi.org/10.1007/s00285-018-1323-x},
	year = {2019},
	publisher = {Springer Science and Business Media {LLC}},
	volume = {78},
	number = {6},
	pages = {1681--1711},
	author = {R.G. Plaza},
	title = {Derivation of a bacterial nutrient-taxis system with doubly degenerate cross-diffusion as the parabolic limit of a velocity-jump process},
	journal = {Journal of Mathematical Biology},
}

@article{Hunt2016,
	doi = {10.1007/s10013-016-0223-x},
	url = {https://doi.org/10.1007/s10013-016-0223-x},
	year = {2016},
	publisher = {Springer Science and Business Media {LLC}},
	volume = {45},
	number = {1-2},
	pages = {221--240},
	author = {A. Hunt and C. Surulescu},
	title = {A Multiscale Modeling Approach to Glioma Invasion with Therapy},
	journal = {Vietnam Journal of Mathematics}
}

@article{Jbabdi2005,
	doi = {10.1002/mrm.20625},
	url = {https://doi.org/10.1002/mrm.20625},
	year = {2005},
	publisher = {Wiley},
	volume = {54},
	number = {3},
	pages = {616--624},
	author = {S. Jbabdi and E. Mandonnet and H. Duffau and L. Capelle and K.R. Swanson and M. P{\'{e}}l{\'{e}}grini-Issac and R. Guillevin and H. Benali},
	title = {Simulation of anisotropic growth of low-grade gliomas using diffusion tensor imaging},
	journal = {Magnetic Resonance in Medicine}
}

@article{swanson2011,
	doi = {10.1158/0008-5472.can-11-1399},
	url = {https://doi.org/10.1158/0008-5472.can-11-1399},
	year = {2011},
	publisher = {American Association for Cancer Research ({AACR})},
	volume = {71},
	number = {24},
	pages = {7366--7375},
	author = {K.R. Swanson and R.C. Rockne and J. Claridge and M.A. Chaplain and E.C. Alvord and A.R.A. Anderson},
	title = {Quantifying the Role of Angiogenesis in Malignant Progression of Gliomas: In Silico Modeling Integrates Imaging and Histology},
	journal = {Cancer Research}
}

@article{hillen2006m5,
  title={M5 mesoscopic and macroscopic models for mesenchymal motion},
  author={Hillen, T.},
  journal={Journal of mathematical biology},
  volume={53},
  number={4},
  pages={585--616},
  year={2006},
  publisher={Springer}
}

@article{conte2020,
	doi = {10.1016/j.jtbi.2019.110088},
	url = {https://doi.org/10.1016/j.jtbi.2019.110088},
	year = {2020},
	publisher = {Elsevier {BV}},
	volume = {486},
	pages = {110088},
	author = {M. Conte and L. Gerardo-Giorda and M. Groppi},
	title = {Glioma invasion and its interplay with nervous tissue and therapy: A multiscale model},
	journal = {Journal of Theoretical Biology}
}

@article{Corbin2018,
	doi = {10.1142/s0218202518400055},
	url = {https://doi.org/10.1142/s0218202518400055},
	year = {2018},
	publisher = {World Scientific Pub Co Pte Lt},
	volume = {28},
	number = {09},
	pages = {1771--1800},
	author = {G. Corbin and A. Hunt and A. Klar and F. Schneider and C. Surulescu},
	title = {Higher-order models for glioma invasion: From a two-scale description to effective equations for mass density and momentum},
	journal = {Mathematical Models and Methods in Applied Sciences}
}

@article{CEKNSSW,
	author = {G. Corbin and C. Engwer and A. Klar and J. Nieto and J. Soler and C. Surulescu and M. Wenske},
	title = {Modeling glioma invasion with anisotropy- and hypoxia-triggered motility enhancement: From subcellular dynamics to macroscopic PDEs with multiple taxis},
	journal = {Mathematical Models and Methods in Applied Sciences},
	volume = {31},
	number = {01},
	pages = {177--222},
	year = {2021},
	doi = {10.1142/S0218202521500056},
}

@article{hanahan2011,
	title={Hallmarks of cancer: the next generation},
	author={D. Hanahan and R.A. Weinberg},
	journal={Cell},
	volume={144},
	number={5},
	pages={646--674},
	year={2011},
}

@article{KELKEL2012,
	doi = {10.1142/s0218202511500175},
	url = {https://doi.org/10.1142/s0218202511500175},
	year = {2012},
	publisher = {World Scientific Pub Co Pte Lt},
	volume = {22},
	number = {03},
	pages = {1150017},
	author = {J. Kelkel and C. Surulescu},
	title = {A multiscale approach to cell migration in tissue networks},
	journal = {Mathematical Models and Methods in Applied Sciences}
}

@article{Lorenz2014,
	doi = {10.1142/s0218202514500249},
	url = {https://doi.org/10.1142/s0218202514500249},
	year = {2014},
	publisher = {World Scientific Pub Co Pte Lt},
	volume = {24},
	number = {12},
	pages = {2383--2436},
	author = {T. Lorenz and C. Surulescu},
	title = {On a class of multiscale cancer cell migration models: Well-posedness in less regular function spaces},
	journal = {Mathematical Models and Methods in Applied Sciences}
}

@article{Kumar20,
	doi = {10.1007/s00285-021-01599-x},
	url = {https://doi.org/10.1007/s00285-021-01599-x},
	year = {2021},
	publisher = {Springer Science and Business Media {LLC}},
	volume = {82},
	number = {6},
	author = {P. Kumar and J. Li and C. Surulescu},
	title = {Multiscale modeling of glioma pseudopalisades: contributions from the tumor microenvironment},
	journal = {Journal of Mathematical Biology}
}

@article{Sidani,
	doi = {10.1083/jcb.200707009},
	url = {https://doi.org/10.1083/jcb.200707009},
	year = {2007},
	publisher = {Rockefeller University Press},
	volume = {179},
	number = {4},
	pages = {777--791},
	author = {M. Sidani and D. Wessels and G. Mouneimne and M. Ghosh and S. Goswami and C. Sarmiento and W. Wang and S. Kuhl and M. El-Sibai and J-M. Backer and R. Eddy and D. Soll and J. Condeelis},
	title = {Cofilin determines the migration behavior and turning frequency of metastatic cancer cells},
	journal = {Journal of Cell Biology},
}

@BOOK{Lauffenburger,
	AUTHOR = {D.A. Lauffenburger and J.L. Lindermann},
	YEAR = {1993},
	Publisher = {Oxford University Press},
	TITLE = {Receptors. Models for binding, trafficing and signaling.},
}

@article{levine2001,
	doi = {10.1006/bulm.2001.0240},
	url = {https://doi.org/10.1006/bulm.2001.0240},
	year = {2001},
	publisher = {Springer Science and Business Media {LLC}},
	volume = {63},
	number = {5},
	pages = {801--863},
	author = {H.A. Levine and Pamuk, S. and Sleeman, B.D. and Nilsen-Hamilton, M.},
	title = {Mathematical Modeling of Capillary Formation and Development in Tumor Angiogenesis: Penetration into the Stroma},
	journal = {Bulletin of Mathematical Biology},
}

@article{Czirok2013,
	doi = {10.1002/wsbm.1233},
	url = {https://doi.org/10.1002/wsbm.1233},
	year = {2013},
	publisher = {Wiley},
	volume = {5},
	number = {5},
	pages = {587--602},
	author = {A. Czirok},
	title = {Endothelial cell motility,  coordination and pattern formation during vasculogenesis},
	journal = {Wiley Interdisciplinary Reviews: Systems Biology and Medicine},
}

@article{Szabo2010,
        doi = {10.1088/1478-3975/7/4/046007},
        url = {https://iopscience.iop.org/article/10.1088/1478-3975/7/4/046007}, 
         year = {2010},
	publisher = {IOP Publishing Ltd},
	volume = {7},
	number = {4},
	pages = {046007},
        author = {A. Szabó and R. \:{U}nnep and E. Méhes and W.O. Twal and W.S. Argraves and Y. Cao and A. Czirók},
        title = {Collective cell motion in endothelial monolayers.},
	journal = {Physical biology}
}

@article{Popel2006,
         year = {2006},
	volume = {2},
	number = {12},
	pages = {e180},
        author = {F. Mac Gabhann and A.S. Popel},
        title = {Targeting neuropilin-1 to inhibit VEGF signaling in cancer: comparison of therapeutic approaches.},
	journal = {PLoS computational biology}
}

@article{gevertz2006,
         year = {2006},
	volume = {243},
	number = {4},
	pages = {517-531},
        author = {J.L. Gevertz and S. Torquato},
        title = {Modeling the effects of vasculature evolution on early brain tumor growth.},
	journal = {Journal of Theoretical Biology}
}

@article{Takano,
         year = {1996},
	volume = {56},
	number = {9},
	pages = {2185-2190},
        author = {S. Takano and Y. Yoshii and S. Kondo and H. Suzuki and T. Maruno and S. Shirai and T. Nose},
        title = {Concentration of vascular endothelial growth factor in the serum and tumor tissue of brain tumor patients.},
	journal = {Cancer research}
}

@article{conte_surulescu2020,
    title={Mathematical modeling of glioma invasion: acid- and vasculature mediated go-or-grow dichotomy and the influence of tissue anisotropy},
    author={M. Conte and C. Surulescu},
    journal = {Applied Mathematics and Computation},
    year={2021},
    volume = {407},
    pages = {126305},
}

@article{qi2006,
  title={An estimation of radiobiologic parameters from clinical outcomes for radiation treatment planning of brain tumor},
  author={X. S. Qi and C.J. Schultz and X.A. Li},
  journal={International Journal of Radiation Oncology, Biology, Physics},
  volume={64},
  number={5},
  pages={1570--1580},
  year={2006},
  publisher={Elsevier}
}

@article{barazzuol2010,
  title={A mathematical model of brain tumour response to radiotherapy and chemotherapy considering radiobiological aspects},
  author={L. Barazzuol and N.G. Burnet and R. Jena and B. Jones and S.J. Jefferies and N.F. Kirkby},
  journal={Journal of theoretical biology},
  volume={262},
  number={3},
  pages={553--565},
  year={2010},
  publisher={Elsevier}
}

@article{kroos2019,
  title={{SDE}-driven modeling of phenotypically heterogeneous tumors: The influence of cancer cell stemness},
  author={J.M. Kroos and C. Stinner and C. Surulescu and N. Surulescu},
  journal={Discrete and Continuous Dynamical Systems - Series B},
  volume={24},
  number={8},
  pages={4629},
  year={2019},
  publisher={American Institute of Mathematical Sciences}
}

@book{joiner2009,
  title={Basic clinical radiobiology fourth edition},
  author={M.C. Joiner and A. Van der Kogel},
  year={2009},
  publisher={CRC press}
}

@article{diao2019,
  title={Behaviors of glioblastoma cells in in vitro microenvironments},
  author={W. Diao and X. Tong and C. Yang and F. Zhang and C. Bao and H. Chen and L. Liu and M. Li and F. Ye and Q. Fan and J. Wang},
  journal={Scientific reports},
  volume={9},
  number={1},
  pages={1--9},
  year={2019},
  publisher={Nature Publishing Group}
}

@article{ke2000,
  title={The relevance of cell proliferation, vascular endothelial growth factor, and basic fibroblast growth factor production to angiogenesis and tumorigenicity in human glioma cell lines},
  author={L.D. Ke and Y.X. Shi and S.A. Im and X. Chen and W.K.A. Yung},
  journal={Clinical Cancer Research},
  volume={6},
  number={6},
  pages={2562--2572},
  year={2000},
  publisher={AACR}
}

@article{kumar2020flux,
	title={A flux-limited model for glioma patterning with hypoxia-induced angiogenesis},
	author={P. Kumar and C. Surulescu},
	journal={Symmetry},
	volume={12},
	number={11},
	pages={1870},
	year={2020},
	publisher={Multidisciplinary Digital Publishing Institute}
}

@misc{TCs_dim,
	howpublished = {https://bionumbers.hms.harvard.edu/bionumber.aspx{?}s=n{\&}v=0{\&}id=108941},
	title = {Estimation taken from},
}

@article{powathil2007,
  title={Mathematical modeling of brain tumors: effects of radiotherapy and chemotherapy},
  author={G. Powathil and M. Kohandel and S. Sivaloganathan and A. Oza and M. Milosevic},
  journal={Physics in Medicine \& Biology},
  volume={52},
  number={11},
  pages={3291},
  year={2007},
  publisher={IOP Publishing},
  doi={10.1088/0031-9155/52/11/023}
}

@misc{ECs_dim,
	howpublished = {http://www.lab.anhb.uwa.edu.au/mb140/MoreAbout/Endothel.htm},
	title = {Estimation taken from},
}

@article{ferrer2018,
  title={Glioma infiltration and extracellular matrix: key players and modulators},
  author={V. Pereira Ferrer and N. Vivaldo Moura and R. Mentlein},
  journal={Glia},
  volume={66},
  number={8},
  pages={1542--1565},
  year={2018},
  publisher={Wiley Online Library}
}

@article{onishi2011,
  title={Angiogenesis and invasion in glioma},
  author={M. Onishi and T. Ichikawa and K. Kurozumi and I. Date},
  journal={Brain tumor pathology},
  volume={28},
  number={1},
  pages={13--24},
  year={2011},
  publisher={Springer}
}

@article{fowler1989,
  title={The linear-quadratic formula and progress in fractionated radiotherapy},
  author={J.F. Fowler},
  journal={The British journal of radiology},
  volume={62},
  number={740},
  pages={679--694},
  year={1989},
  publisher={The British Institute of Radiology}
}

@article{vempati2014,
  title={Extracellular regulation of VEGF: isoforms, proteolysis, and vascular patterning},
  author={P. Vempati and A.S. Popel and F. Mac Gabhann},
  journal={Cytokine \& growth factor reviews},
  volume={25},
  number={1},
  pages={1--19},
  year={2014},
  publisher={Elsevier}
}

@article{dietrich2020,
  title={Multiscale modeling of glioma invasion: from receptor binding to flux-limited macroscopic {PDE}s},
  author={A. Dietrich and N. Kolbe and N. Sfakianakis and C. Surulescu},
  journal={Multiscale Modeling and Simulation},
  year={2022},
  volume={20},
  number={2},
  pages={685-713},
}

@book{seyfried2012,
  title={Cancer as a metabolic disease: on the origin, management, and prevention of cancer},
  author={T. Seyfried},
  year={2012},
  publisher={John Wiley \& Sons}
}

@article{stupp2005,
  title={Radiotherapy plus concomitant and adjuvant temozolomide for glioblastoma},
  author={R. Stupp and W.P. Mason and M.J. Van Den Bent and M. Weller and B. Fisher and M.J.B. Taphoorn and K. Belanger and A.A. Brandes and C. Marosi and U. Bogdahn and others},
  journal={New England journal of medicine},
  volume={352},
  number={10},
  pages={987--996},
  year={2005},
  publisher={Mass Medical Soc}
}

@article{friedman2009,
  title={Bevacizumab alone and in combination with irinotecan in recurrent glioblastoma},
  author={H.S. Friedman and M.D. Prados and P.Y. Wen and T. Mikkelsen and D. Schiff and L.E. Abrey and W.K.A. Yung and N. Paleologos and M.K. Nicholas and R. Jensen and others},
  journal={Journal of clinical oncology},
  volume={27},
  number={28},
  pages={4733--4740},
  year={2009},
  publisher={American Society of Clinical Oncology}
}

@article{stupp2009,
  title={Effects of radiotherapy with concomitant and adjuvant temozolomide versus radiotherapy alone on survival in glioblastoma in a randomised phase III study: 5-year analysis of the {EORTC-NCIC} trial},
  author={R. Stupp and M.E. Hegi and W.P. Mason and M.J. Van Den Bent and M.J.B. Taphoorn and R.C. Janzer and S.K. Ludwin and A. Allgeier and B. Fisher and K. Belanger and others},
  journal={The lancet oncology},
  volume={10},
  number={5},
  pages={459--466},
  year={2009},
  publisher={Elsevier}
}

@article{knobe2021,
  title={Feasibility and clinical usefulness of modelling glioblastoma migration in adjuvant radiotherapy},
  author={S. Knobe and Y. Dzierma and M. Wenske and C. Berdel and J. Fleckenstein and P. Melchior and J. Palm and F. Nuesken and A. Hunt and C. Engwer and others},
  journal={Zeitschrift f{\"u}r Medizinische Physik},
  year={2021},
  publisher={Elsevier}
}

@article{shapiro99,
	author = {Shapiro, W.R.},
	title = "{Current Therapy for Brain Tumors: Back to the Future}",
	journal = {Archives of Neurology},
	volume = {56},
	number = {4},
	pages = {429-432},
	year = {1999},
	doi = {10.1001/archneur.56.4.429},
	url = {https://doi.org/10.1001/archneur.56.4.429},
}

@article{ZS22,
doi = {10.1137/20m1365442},
url = {https://doi.org/10.1137%2F20m1365442},
year = 2022,
publisher = {Society for Industrial {\&} Applied Mathematics ({SIAM})},
volume = {82},
number = {1},
pages = {142--167},
author = {A. Zhigun and C. Surulescu},
title = {A Novel Derivation of Rigorous Macroscopic Limits from a Micro-Meso Description of Signal-Triggered Cell Migration in Fibrous Environments},
journal = {{SIAM} Journal on Applied Mathematics}
}

@article{Rockne2010,
	doi = {10.1088/0031-9155/55/12/001},
	url = {https://doi.org/10.1088%2F0031-9155%2F55%2F12%2F001},
	year = 2010,
	publisher = {{IOP} Publishing},
	volume = {55},
	number = {12},
	pages = {3271--3285},
	author = {R. Rockne and J.K. Rockhill and M. Mrugala and A.M. Spence and I. Kalet and K. Hendrickson and A. Lai and T. Cloughesy and E.C. Alvord and K.R. Swanson},
	title = {Predicting the efficacy of radiotherapy in individual glioblastoma patients in vivo:a mathematical modeling approach},
	journal = {Physics in Medicine and Biology}
}

@article{Holdsworth_2012,
	doi = {10.1088/0031-9155/57/24/8271},
	url = {https://doi.org/10.1088%2F0031-9155%2F57%2F24%2F8271},
	year = 2012,
	publisher = {{IOP} Publishing},
	volume = {57},
	number = {24},
	pages = {8271--8283},
	author = {C.H. Holdsworth and D. Corwin and R.D. Stewart and R. Rockne and A.D. Trister and K.R. Swanson and M. Phillips},
	title = {Adaptive {IMRT} using a multiobjective evolutionary algorithm integrated with a diffusion{\textendash}invasion model of glioblastoma},
	journal = {Physics in Medicine and Biology}
}

@article{Swan_2017,
	doi = {10.1007/s11538-017-0271-8},
	url = {https://doi.org/10.1007%2Fs11538-017-0271-8},
	year = 2017,
	publisher = {Springer Science and Business Media {LLC}},
	volume = {80},
	number = {5},
	pages = {1259--1291},
	author = {A. Swan and T. Hillen and J.C. Bowman and A.D. Murtha},
	title = {A Patient-Specific Anisotropic Diffusion Model for Brain Tumour Spread},
	journal = {Bulletin of Mathematical Biology}
}

@article{stummer2009extent,
  title={Extent of resection and survival in glioblastoma multiforme},
  author={W. Stummer},
  journal={Neurosurgery},
  volume={64},
  number={6},
  pages={E1206},
  year={2009},
  publisher={Oxford University Press}
}

@article{reardon2011review,
  title={A review of VEGF/VEGFR-targeted therapeutics for recurrent glioblastoma},
  author={D.A. Reardon and S. Turner and K.B. Peters and A. Desjardins and S. Gururangan and J.H. Sampson and R.E. McLendon and J.E. Herndon and L.W. Jones and J.P. Kirkpatrick and others},
  journal={Journal of the National Comprehensive Cancer Network},
  volume={9},
  number={4},
  pages={414--427},
  year={2011},
  publisher={Harborside Press, LLC}
}

\end{document}